\documentclass[12pt]{article}
\usepackage{a4wide}
\usepackage{latexsym}
\usepackage{amsmath}
\usepackage{amssymb}
\usepackage{amsfonts}
\usepackage{amscd}
\usepackage{cite}
\usepackage{graphicx}
\usepackage{axodraw}
\usepackage{float}

\usepackage{pslatex}
\usepackage[latin1]{inputenc}
\usepackage[OT2,T1]{fontenc}

\newcommand{\bq}{\begin{eqnarray}}
\newcommand{\eq}{\end{eqnarray}}

\DeclareSymbolFont{cyrletters}{OT2}{wncyr}{m}{n}
\DeclareMathSymbol{\Sha}{\mathalpha}{cyrletters}{"58}

\begin{document}

\thispagestyle{empty}

\begin{flushright}
  KA-TP-28-2013 \\
  MITP/13-058
\end{flushright}

\vspace{1.5cm}

\begin{center}
  {\Large\bf Decomposition of one-loop QCD amplitudes into primitive amplitudes based on shuffle relations\\
  }
  \vspace{1cm}
  {\large Christian Reuschle ${}^{a}$ and Stefan Weinzierl ${}^{b}$\\
  \vspace{4mm}
      {\small ${}^{a}$ \em Institute for Theoretical Physics, }
      {\small \em Karlsruhe Institute of Technology,}\\
      {\small \em D - 76128 Karlsruhe, Germany}\\
  \vspace{4mm}
      {\small ${}^{b}$ \em PRISMA Cluster of Excellence, Institut f{\"u}r Physik, }\\
      {\small \em Johannes Gutenberg-Universit{\"a}t Mainz,}
      {\small \em D - 55099 Mainz, Germany}\\
  \vspace{4mm}
  } 
\end{center}

\vspace{2cm}

\begin{abstract}\noindent
  {
We present the decomposition of QCD partial amplitudes into primitive amplitudes at one-loop level and tree level
for arbitrary numbers of quarks and gluons.
Our method is based on shuffle relations.
This method is purely combinatorial and does not require the inversion of a system of linear equations.
   }
\end{abstract}

\vspace*{\fill}

\newpage

\section{Introduction}
\label{sect:intro}

The recent years have witnessed a significant advance in our abilities to compute QCD NLO corrections
and allowed for the computation of multi-parton observables at NLO level.
New methods like the 
unitarity method \cite{Bern:1995cg,Britto:2004nc,Forde:2007mi,Ossola:2006us,Mastrolia:2008jb,Anastasiou:2006jv,Anastasiou:2006gt,Ellis:2007br,Giele:2008ve,Ellis:2008ir}
or numerical approaches \cite{Assadsolimani:2009cz,Assadsolimani:2010ka,Becker:2010ng,Becker:2011vg,Becker:2012aq,Gotz:2012zz,Becker:2012nk,Becker:2012bi}
opened the door to multi-parton final states.
These methods organise the computation of the one-loop amplitude as a sum over smaller pieces, called primitive amplitudes.
The most important features of a primitive amplitude are gauge invariance and a fixed cyclic ordering of the external legs.
(A more precise definition will be given later on.)
Primitive amplitudes should not be confused with partial amplitudes, which are the kinematic coefficients of the independent colour
structures.
Partial amplitudes are also gauge invariant, but not necessarily cyclic ordered.
The leading contributions in an $1/N$-expansion (with $N$ being the number of colours)
are usually cyclic ordered, the sub-leading parts are in general not.
The decomposition of the full one-loop amplitude into partial amplitudes is easily derived.
However, it is less trivial to find a decomposition of the partial amplitudes into primitive amplitudes.
A closed formula is known in the special cases of the $n$-gluon amplitudes \cite{Bern:1990ux}
and the amplitudes with one quark-antiquark pair plus $(n-2)$ gluons \cite{Bern:1994fz}.
It should be mentioned that the decomposition of partial amplitudes into primitive amplitudes is in general not unique.

For amplitudes with more than one quark-antiquark pair an algorithm is known \cite{Ellis:2008qc,Ellis:2011cr,Ita:2011ar,Badger:2012pg}, 
which expresses partial amplitudes in terms of primitive amplitudes.
This algorithm is based on Feynman diagrams and the solution of a system of linear equations.
From a pragmatic point of view, the above mentioned algorithm solves the problem.
However, it is unsatisfactory for the following two reasons:
First of all one has to solve a (large) system of linear equations. A method which avoids this is certainly preferred.
Secondly, the method relies on Feynman diagrams. This is unaesthetic, as all other parts of a next-to-leading order calculation can be performed
without resorting to Feynman diagrams.
There is no compelling reason why Feynman diagrams should be needed in the decomposition of partial amplitudes into primitive amplitudes.
It only reflects the fact, that up to now we do not know a better method.

In this paper we examine in more detail the decomposition of partial amplitudes into primitive amplitudes.
We present a method, which expresses partial amplitudes as a linear combination of primitive amplitudes.
Our method avoids Feynman diagrams and the inversion of a system of linear equations.
Instead within our approach the decomposition is given by (generalised) shuffle relations.
They generalise the known decompositions for the all-gluon case and the quark-antiquark plus $(n-2)$-gluons case to all QCD amplitudes.

On the side we would like to mention a few papers not directly related to our work, but relevant in the wider context of the colour
decomposition of QCD amplitudes: These approaches are the decomposition into an orthogonal base of irreducible $SU(N)$-multiplets  \cite{Keppeler:2012ih,Zeppenfeld:1988bz},
the colour decomposition in a spontaneously broken gauge theory \cite{Dai:2012jh} 
and a formal treatment of the $U(1)$-gluons \cite{Kilian:2012pz}.

This paper is organised as follows:
In section~\ref{sect:math} we review the required mathematical background on permutations and shuffle algebras.
In section~\ref{sect:amplitudes} the known facts on the colour decomposition of one-loop (and tree-level) amplitudes are
summarised.
Our method is based on a few basic shuffle operations, which are presented in section~\ref{sect:operations}.
The decomposition of the partial amplitudes in terms of primitive amplitudes is presented in section~\ref{sect:method}.
Section~\ref{sect:examples} illustrates our method with a few examples.
Finally, our conclusions are given in section~\ref{sect:conclusions}.
In an appendix we have collected the cyclic ordered Feynman rules (appendix~\ref{appendix:colour_ordered_rules})
as well as proofs on shuffle relations (appendices~\ref{appendix:U1_shuffle} and~\ref{appendix:loop_closing}),
which are too technical to be included in the main text.

\section{Mathematical preliminaries}
\label{sect:math}

In this section we review the basics of permutations and shuffle algebras as far as they are needed in sequel of the paper.

\subsection{Permutations}

Let $A=\{\alpha_1,...,\alpha_n\}$ be a set of $n$ elements.
A permutation $\sigma$ is a bijective map
\bq
 \sigma & : & A \rightarrow A.
\eq
For a given set $A$ the set of all permutations $\sigma$ forms a group, called the symmetric group $S_n(A)$.
If $A=\{1,...,n\}$, we simply write $S_n$.
Repeated application of $\sigma$ yields again a permutation and we write 
\bq
 \sigma^k & = & \underbrace{\sigma \circ ... \circ \sigma}_{k \; \mathrm{times}}.
\eq
An element $\alpha_j \in A$ is called of order $k$, if
\bq
 \sigma^k\left(\alpha_j\right) & = & \alpha_j,
\eq
and
\bq
 \sigma^m\left(\alpha_j\right) & \neq & \alpha_j 
 \;\;\;\;\;\; \mbox{for} \;\; m < k.
\eq
In this case the ordered sequence $(\sigma^0(\alpha_j), \sigma^1(\alpha_j), ..., \sigma^{k-1}(\alpha_j))$  forms a cycle.
Permutations may be denoted as products of cycles. For example, the permutation given by the two-line notation on the left-hand side
\bq
 \left( \begin{array}{ccccc}
 1 & 2 & 3 & 4 & 5 \\
 3 & 4 & 5 & 2 & 1 \\
 \end{array} \right)
 & = &
 \left( 1, 3, 5 \right) \left( 2, 4 \right)
\eq
is identical to the one defined by the cycle notation on the right-hand side
with the two cycles $(1,3,5)$ and $(2,4)$.

\subsection{Shuffle algebras}
\label{subsection:shuffle}

Consider a set of letters $A$. 
The set $A$ is called the alphabet.
A word is an ordered sequence of letters $l_i \in A$:
\bq
 w & = & l_1 l_2 ... l_k.
\eq
The word of length zero is denoted by $e$.
Let $K$ be a field and consider the vector space of words over $K$.
A shuffle algebra ${\cal A}$ on the vector space of words is defined by
the shuffle product $\Sha$ through
\bq
 l_1 l_2 ... l_k \; \Sha \; l_{k+1} ... l_r 
 & = &
 \sum\limits_{\mbox{\tiny shuffles} \; \sigma} l_{\sigma(1)} l_{\sigma(2)} ... l_{\sigma(r)},
\eq
where the sum runs over all permutations $\sigma$, which preserve the relative order of $l_1,l_2,...,l_k$ and of $l_{k+1},...,l_r$.
The shuffle product is commutative and associative.
The name ``shuffle algebra'' is related to the analogy of shuffling cards: If a deck of cards
is split into two parts and then shuffled, the relative order within the two individual parts
is conserved. 
The name ``ordered permutations'' is also used for the shuffle product.
The empty word $e$ is the unit in this algebra:
\bq
 e \cdot w = w \cdot e = w.
\eq
For a word $w=l_1 l_2 ... l_k$ we introduce a left-shift operator $L$ and a right-shift operator $R$
by
\bq
 L \left( l_1 l_2 ... l_k \right) = l_2 ... l_k l_1,
 & &
 R \left( l_1 ... l_{k-1} l_k \right) = l_k l_1 ... l_{k-1}.
\eq
It is often the case, that we are only interested in an ordered sequence up to cyclic permutations, in other words we consider
two words to be equivalent, if they differ only by a cyclic permutation:
\bq
 l_1 l_2 ... l_k & \sim & l_2 ... l_k l_1.
\eq
We call the equivalence classes ``cyclic ordered words'' and denote these by brackets $()$ around a representative word.
For cyclic ordered words we can define a cyclic shuffle product by
\bq
\left( l_1 l_2 ... l_k \right) \circledcirc
 \left( l_{k+1} ... l_r \right) & = &
 \sum\limits_{(\mbox{\tiny cyclic shuffles} \; \sigma) / {\mathbb Z}_r} \left( l_{\sigma(1)} l_{\sigma(2)} ... l_{\sigma(r)} \right),
\eq
where the sum runs over all permutations $\sigma$, which preserve the relative cyclic order of $l_1,l_2,...,l_k$ and of $l_{k+1},...,l_r$
modulo the cyclic permutations ${\mathbb Z}_r$.
The cyclic shuffle product is commutative and associative.
The name ``cyclic ordered permutations'' is also used for the cyclic shuffle product.
Two examples illustrate the differences between the shuffle product and the cyclic shuffle product:
\bq
 l_1 l_2 \; \Sha \; l_3
 & = &
 l_1 l_2 l_3 + l_1 l_3 l_2 + l_3 l_1 l_2,
 \nonumber \\
 \left( l_1 l_2 \right) \circledcirc \left( l_3 \right)
 & = &
 \left( l_1 l_2 l_3 \right) + \left( l_1 l_3 l_2 \right).
\eq
In general we have the following relation between the cyclic shuffle product ``$\circledcirc$'' and the shuffle product ``$\Sha$'':
\bq
\left( l_1 l_2 ... l_k \right) \circledcirc
 \left( l_{k+1} ... l_r \right) & = &
 \sum\limits_{j=0}^{r-k-1} \left( l_1 \left( l_2 ... l_k \; \Sha \; L^j\left( l_{k+1} ... l_r \right) \right) \right).
\eq
The right-hand side gives all cyclic ordered words with letter $l_1$ in front and which preserve the relative cyclic order of 
$l_1,l_2,...,l_k$ and of $l_{k+1},...,l_r$.
In a similar way we can always take the representative, where the letter $l_r$ appears in the end. Thus we have equivalently
\bq
\left( l_1 l_2 ... l_k \right) \circledcirc
 \left( l_{k+1} ... l_r \right) & = &
 \sum\limits_{j=0}^{k-1} \left( \left( L^j\left( l_1 l_2 ... l_k \right) \; \Sha \; l_{k+1} ... l_{r-1} \right) l_r \right).
\eq
We will apply the shuffle product and the cyclic shuffle product to strings made out of
the alphabet $A=\{\bar{q}_1,\bar{q}_2,...,q_1,q_2,...,g_1,g_2,...\}$, where $\bar{q}_i$ corresponds to an external antiquark, $q_j$ to an
external quark and $g_k$ to an external gluon.
We are in particular interested in (cyclic) orderings, which correspond to amplitudes, whose Feynman diagrams can be drawn in a planar way
on a disc (for tree amplitudes) or on an annulus (for loop amplitudes).
This excludes in particular crossed fermion lines.
\begin{figure}
\begin{center}
\begin{picture}(110,100)(0,0)
 \CArc(50,50)(40,0,360)
 \ArrowLine(21.7,21.7)(21.7,78.3)
 \ArrowLine(78.3,78.3)(78.3,21.7)
 \Text(21.7,16)[t]{$\bar{q}_1$}
 \Text(21.7,84)[b]{$q_1$}
 \Text(78.3,84)[b]{$\bar{q}_2$}
 \Text(78.3,16)[t]{$q_2$}
\end{picture} 
\begin{picture}(110,100)(0,0)
 \CArc(50,50)(40,0,360)
 \Line(21.7,21.7)(50,50)
 \ArrowLine(50,50)(78.3,78.3)
 \Line(21.7,78.3)(50,50)
 \ArrowLine(50,50)(78.3,21.7)
 \Text(21.7,16)[t]{$\bar{q}_1$}
 \Text(21.7,84)[b]{$\bar{q}_2$}
 \Text(78.3,84)[b]{$q_1$}
 \Text(78.3,16)[t]{$q_2$}
\end{picture} 
\begin{picture}(110,100)(0,0)
 \CArc(50,50)(40,0,360)
 \CArc(50,50)(10,0,360)
 \ArrowLine(21.7,21.7)(21.7,78.3)
 \ArrowLine(78.3,78.3)(78.3,21.7)
 \Text(21.7,16)[t]{$\bar{q}_1^L$}
 \Text(21.7,84)[b]{$q_1^L$}
 \Text(78.3,84)[b]{$\bar{q}_2^L$}
 \Text(78.3,16)[t]{$q_2^L$}
\end{picture} 
\begin{picture}(100,100)(0,0)
 \CArc(50,50)(40,0,360)
 \CArc(50,50)(10,0,360)
 \Line(21.7,21.7)(60,30)
 \ArrowArc(60,50)(20,-90,90)
 \Line(60,70)(21.7,78.3)
 \Line(78.3,78.3)(40,70)
 \ArrowArc(40,50)(20,90,270)
 \Line(40,30)(78.3,21.7)
 \Text(21.7,16)[t]{$\bar{q}_1^R$}
 \Text(21.7,84)[b]{$q_1^R$}
 \Text(78.3,84)[b]{$\bar{q}_2^R$}
 \Text(78.3,16)[t]{$q_2^R$}
\end{picture} 
\caption{\label{figure_no_crossed_fermion_lines}
Illustration of planar and non-planar cyclic orderings: The cyclic ordering $(\bar{q}_1 q_1 \bar{q}_2 q_2)$ can be drawn in a planar way on a disc 
(left diagram), while the cyclic ordering $(\bar{q}_1 \bar{q}_2 q_1 q_2)$ cannot be drawn in a planar way on a disc (second-to-left diagram).
The cyclic ordering $(\bar{q}_1^L q_1^L \bar{q}_2^L q_2^L)$ with left/right assignments can be drawn in a planar way on an annulus (second-to-right diagram),
while the cyclic ordering $(\bar{q}_1^R q_1^R \bar{q}_2^R q_2^R)$ with left/right assignments cannot be drawn in a planar way on an annulus (right diagram). 
}
\end{center}
\end{figure}
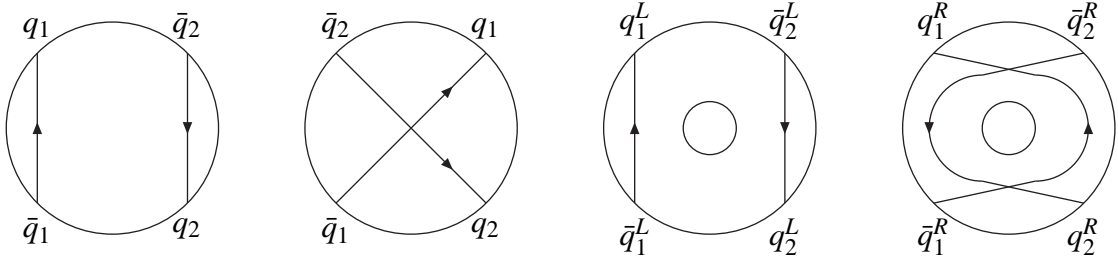
We consider first tree amplitudes.
Formally, we can define a projection operator $P_{\mathrm{no \; crossed \; fermions}}$, which projects onto (cyclic) words corresponding to no crossed fermion lines.
In an ordered sequence we can take the first appearance of a quark or an antiquark of flavour $i$ as an opening bracket of type $i$, and the
appearance of the corresponding antiquark or quark as a closing bracket of type $i$.
Closing brackets of type $i$ only match with opening brackets of type $i$.
We then say that a word has no crossed fermion lines, if it corresponds to properly matched brackets.
This is a generalisation of Dyck words \cite{Melia:2013bta}. For Dyck words one does not distinguish between brackets of different types.
For a cyclic ordered sequence we say that the cyclic word has no crossed fermion lines, if there is a cyclic permutation such that
the corresponding representative is an ordered sequence with no crossed fermion lines.
This is illustrated in fig.~(\ref{figure_no_crossed_fermion_lines}).

We then define $P_{\mathrm{no \; crossed \; fermions}}$ as the projection onto (cyclic) words with no crossed fermions.
For example,
\bq
 P_{\mathrm{no \; crossed \; fermions}}\left( \bar{q}_1 q_1 \bar{q}_2 q_2 \right) & = & \bar{q}_1 q_1 \bar{q}_2 q_2,
 \nonumber \\
 P_{\mathrm{no \; crossed \; fermions}}\left( \bar{q}_1 \bar{q}_2 q_1 q_2 \right) & = & 0.
\eq
In loop amplitudes the quarks and antiquarks carry an additional label ``L'' or ``R'', depending on whether
the fermion line passes to the left or to the right of the loop.
When we draw the corresponding diagrams on an annulus, such that the fermion lines start and end on the outer boundary, we obtain additional
restrictions, if we require that fermion lines do not cross. For example
\bq
 P_{\mathrm{no \; crossed \; fermions}}\left( \bar{q}_1^L q_1^L \bar{q}_2^L q_2^L \right) & = & \bar{q}_1^L q_1^L \bar{q}_2^L q_2^L,
 \nonumber \\
 P_{\mathrm{no \; crossed \; fermions}}\left( \bar{q}_1^R q_1^R \bar{q}_2^L q_2^L \right) & = & \bar{q}_1^R q_1^R \bar{q}_2^L q_2^L,
 \nonumber \\
 P_{\mathrm{no \; crossed \; fermions}}\left( \bar{q}_1^L q_1^L \bar{q}_2^R q_2^R \right) & = & \bar{q}_1^L q_1^L \bar{q}_2^R q_2^R,
\eq
but
\bq
 P_{\mathrm{no \; crossed \; fermions}}\left( \bar{q}_1^R q_1^R \bar{q}_2^R q_2^R \right) & = & 0.
\eq
Two examples are illustrated in fig.~(\ref{figure_no_crossed_fermion_lines}).
We can define a modified shuffle product and a modified cyclic shuffle product by
\bq
\label{shuffle_no_crossed_fermions}
  P_{\mathrm{no \; crossed \; fermions}}\left( l_1 l_2 ... l_k \; \Sha \; l_{k+1} ... l_r \right),
 \nonumber \\
  P_{\mathrm{no \; crossed \; fermions}}\left( \left( l_1 l_2 ... l_k \right) \circledcirc \left( l_{k+1} ... l_r \right) \right),
\eq
where $l_1 l_2 ... l_k$ and $l_{k+1} ... l_r$ are words with no crossed fermion lines and 
$(l_1 l_2 ... l_k)$ and $(l_{k+1} ... l_r)$ are cyclic words with no crossed fermion lines.
The modified products are again commutative and associative.
By abuse of notation we will denote these products again by ``$\Sha$'' and ``$\circledcirc$''.
We do not distinguish between the unmodified and modified products for the following reason: 
We will later associate to cyclic words planar amplitudes. Cyclic words with crossed fermions correspond to planar amplitudes
with crossed fermions (therefore the name).
The Feynman diagrams corresponding to these amplitudes can only be drawn in a planar way with flavour-changing currents.
However, in QCD there are no flavour-changing currents and these amplitudes are zero.
From now on we will use the convention, that words or cyclic words with crossed fermions are immediately set to zero.

We introduce the following convenient notation: Suppose $f(w)$ is a function of cyclic words $w$ and $S = w_1 \circledcirc w_2$,
like for example
\bq
 S & = & \left(g_1g_2\right) \circledcirc \left(g_3\right)
 = \left(g_1g_2g_3\right) + \left(g_1g_3g_2\right).
\eq
Then we write
\bq
 \sum\limits_{w\in w_1 \circledcirc w_2} f\left(w\right)
\eq
for
\bq
 f\left(g_1g_2g_3\right) + f\left(g_1g_3g_2\right).
\eq
If $\lambda_1, \lambda_2$ are numbers and $w_1, w_2$ cyclic words, we write
\bq
 \sum\limits_{w \in \lambda_1 w_1 + \lambda_2 w_2} f\left(w\right)
\eq
for
\bq
 \lambda_1 f\left(w_1\right)
 +
 \lambda_2 f\left(w_2\right).
\eq
In other words, we take $f$ as a linear operator on the vector space of cyclic words.

\section{Colour decomposition of amplitudes}
\label{sect:amplitudes}

In this section we review known facts on the colour decomposition of amplitudes.
We first show how to reduce amplitudes with quarks of identical flavour to the non-identical case.
We then define partial amplitudes and primitive amplitudes.
We further outline the known algorithm for the decomposition of partial amplitudes into primitive amplitudes based on Feynman
diagrams and linear equations.

\subsection{Amplitudes with identical quarks}

Amplitudes with quark-antiquark pairs of identical flavour can always be related to amplitudes,
where all quark-antiquark pairs have different flavours.
This is achieved by summing over all quark permutations.
An amplitude with $n_q$ quark-antiquark pairs can be written as
\bq
\label{identical_quarks}
\lefteqn{
{\cal A}\left( \bar{q}_1, q_1, ..., \bar{q}_2, q_2, ..., \bar{q}_{n_q}, q_{n_q} \right)
 = } & & \nonumber \\
 & & 
 \sum\limits_{\sigma \in S(n_q)} \left( -1 \right)^{\sigma} 
 \left( \prod\limits_{j=1}^{n_q} \delta^{\mathrm{flav}}_{\bar{q}_j q_{\sigma(j)} } \right)
     \hat{\cal A}\left( \bar{q}_1, q_{\sigma(1)}, ..., \bar{q}_2, q_{\sigma(2)}, ..., \bar{q}_{n_q}, q_{\sigma(n_q)} \right).
\eq
Here, $(-1)^\sigma$ equals $-1$ whenever the permutation is odd and equals $+1$ if the permutation is even.
In $\hat{\cal A}$ each external quark-antiquark pair $(\bar{q}_j, q_{\sigma(j)})$ is connected by a continuous fermion line.
The flavour factor $\delta^{flav}_{\bar{q}_j q_{\sigma(j)} }$ ensures that this combination is only taken into account, if
$\bar{q}_j$ and $ q_{\sigma(j)}$ have the same flavour.
In $\hat{\cal A}$ each external quark-antiquark pair $(\bar{q}_j, q_{\sigma(j)})$ is treated as having a flavour different from all other
quark-antiquark pairs.
It is therefore sufficient to discuss only the case of different quark flavours 
and we will therefore from now on assume that all quark flavours are different.

\subsection{Partial amplitudes}

Amplitudes in QCD may be decomposed into group-theoretical factors 
(carrying the colour structures) multiplied by kinematic functions called partial amplitudes
\cite{Cvitanovic:1980bu,Berends:1987cv,Mangano:1987xk,Kosower:1987ic,Bern:1990ux}. 
These partial amplitudes do not contain any colour information and are gauge invariant objects. 
The colour decomposition is obtained by replacing the structure constants $f^{abc}$ by
\bq
 i f^{abc} & = & 2 \left[ \mbox{Tr}\left(T^a T^b T^c\right) - \mbox{Tr}\left(T^b T^a T^c\right) \right],
\eq
which follows from $ \left[ T^a, T^b \right] = i f^{abc} T^c$. In this paper we use the normalisation
\bq
 \mbox{Tr}\;T^a T^b & = & \frac{1}{2} \delta^{a b}
\eq
for the colour matrices. The resulting traces and strings of colour matrices 
can be further simplified with the help of the Fierz identity :
\bq
\label{fierz_identity}
 T^a_{ij} T^a_{kl} & = &  \frac{1}{2} \left( \delta_{il} \delta_{jk}
                         - \frac{1}{N} \delta_{ij} \delta_{kl} \right).
\eq
There are several possible choices for a basis in colour space. 
A convenient choice is the colour-flow basis \cite{'tHooft:1973jz,Maltoni:2002mq,Weinzierl:2005dd}. 
This choice is obtained by attaching a factor
\bq
 \sqrt{2} T^a_{ij}
\eq
to each external gluon and by using subsequently the Fierz identity in eq.~(\ref{fierz_identity})
to eliminate the adjoint indices.
In squaring these amplitudes a colour projector
\bq
\label{projection_operator}
 \delta_{\bar{i} i} \delta_{j \bar{j}} - \frac{1}{N} \delta_{\bar{i} \bar{j} } \delta_{j i}
\eq
has to be applied to each gluon when one works within the colour-flow basis.

As an example we first consider the colour decomposition of the Born pure gluon amplitude with $n$ external gluons.
The colour decomposition of the amplitude may be written in 
the form
\bq
\label{colour_decomp_pure_gluon}
 {\cal A}_{n}^{(0)}(g_1,g_2,...,g_n) 
 & = & 
 \left(\frac{g}{\sqrt{2}}\right)^{n-2} 
 \sum\limits_{\sigma \in S_{n}/Z_{n}} 
 \delta_{i_{\sigma_1} j_{\sigma_2}} \delta_{i_{\sigma_2} j_{\sigma_3}} 
 ... \delta_{i_{\sigma_n} j_{\sigma_1}}  
 A_{n}^{(0)}\left( g_{\sigma_1}, ..., g_{\sigma_n} \right),
\eq
where the sum is over all non-cyclic permutations of the external gluon legs.
The quantities $A^{(0)}_n(g_{\sigma_1},...,g_{\sigma_n})$, called the partial amplitudes, contain the 
kinematic information.
As a further example we give the colour decomposition 
for a tree amplitude with a pair of quarks:
\bq
\label{colour_decomp_qqbar_gluon}
 {\cal A}_{n}^{(0)}(q,g_1,...,g_{n-2},\bar{q}) 
 & = & 
 \left(\frac{g}{\sqrt{2}}\right)^{n-2} 
 \sum\limits_{\sigma \in S_{n-2}} 
 \delta_{i_q j_{\sigma_1}} \delta_{i_{\sigma_1} j_{\sigma_2}} 
 ... \delta_{i_{\sigma_{n-2}} j_{\bar{q}}} 
A_{n}^{(0)}(q,g_{\sigma_1},...,g_{\sigma_{n-2}},\bar{q}),
\;\;\;\;\;\;\;\;\;
\eq
where the sum is over all permutations of the external gluon legs. 

In these examples we have two basic colour structures, a colour cluster described by the ``closed string''
\bq
 c_{\mathrm{closed}}(g_1,...,g_n) 
 & = &
 \delta_{i_{1} j_{2}} \delta_{i_{2} j_{3}} 
 ... \delta_{i_{n} j_{1}}  
\eq
and a colour cluster corresponding to the ``open string''
\bq
 c_{\mathrm{open}}(q,g_1,...,g_n,\bar{q})
 & = &
 \delta_{i_q j_{1}} \delta_{i_{1} j_{2}} 
 ... \delta_{i_{n} j_{\bar{q}}}.
\eq
Two special cases for the closed string are worth mentioning: The colour structure of a closed string with one gluon 
is simply $c_{\mathrm{closed}}(g_1) = \delta_{i_1j_1}$.
This gives a vanishing contribution to the amplitude squared due to the projection operator in eq.~(\ref{projection_operator}).
If one works in a colour basis consisting of traces and strings of generators in the fundamental representation this corresponds 
to the fact that the generators for $SU(N)$ are traceless
\bq
 \mathrm{Tr}\;T^a & = & 0.
\eq
The second special case is the closed string with no gluon attached. We define the empty closed string to be equal to $c_{\mathrm{closed}}()=N$.
The motivation comes from the fact that in a colour basis consisting of traces and strings of generators in the fundamental representation
we have
\bq
 \mathrm{Tr}\;{\bf 1} & = & N.
\eq
With this notation the colour decomposition of the Born pure gluon amplitude in eq.~(\ref{colour_decomp_pure_gluon}) becomes
\bq
\label{colour_decomp_pure_gluon_II}
 {\cal A}_{n}^{(0)}(g_1,g_2,...,g_n) 
 & = & 
 \left(\frac{g}{\sqrt{2}}\right)^{n-2} 
 \sum\limits_{\sigma \in S_{n}/Z_{n}} 
 c_{\mathrm{closed}}\left( g_{\sigma_1}, g_{\sigma_2}, ..., g_{\sigma_n} \right)
 A_{n}^{(0)}\left( g_{\sigma_1}, ..., g_{\sigma_n} \right),
\eq
Born amplitudes with additional pairs of quarks have a decomposition in colour factors, which are products
of open strings. The colour decomposition of a Born amplitude with $n_q$ quarks, $n_q$ antiquarks and $n_g$ gluons 
(and therefore $n=n_g+2n_q$ external particles) reads \cite{Mangano:1990by,Maltoni:2002mq,Weinzierl:2005dd}
\bq
\label{tree_multi_quark}
 {\cal A}^{(0)}_n & = &
 \left(\frac{g}{\sqrt{2}}\right)^{n-2}
 \sum\limits_{\sigma \in S_{n_g}} \sum\limits_{\pi \in S_{n_q}}
 \sum\limits_{\stackrel{i_1,...,i_{n_q} \ge 0}{i_1+...+i_{n_q}=n_g}}
 c_{\mathrm{open}}\left(q_1,g_{\sigma_1},...,g_{\sigma_{i_1}},\bar{q}_{\pi_1}\right)
 \nonumber \\
 & & 
 c_{\mathrm{open}}\left(q_2,g_{\sigma_{i_1+1}},...,g_{\sigma_{i_1+i_2}},\bar{q}_{\pi_2}\right)
 ...
 c_{\mathrm{open}}\left(q_{n_q},g_{\sigma_{i_1+...+i_{n_q-1}+1}},...,g_{\sigma_{i_1+...+i_{n_q}}},\bar{q}_{\pi_{n_q}}\right)
 \nonumber \\
 & &
 A^{(0)}_n\left(q_1,g_{\sigma_1},...,g_{\sigma_{i_1}},\bar{q}_{\pi_1}, q_2, ..., g_{\sigma_{i_1+...+i_{n_q}}}, \bar{q}_{\pi_{n_q}} \right).
\eq
The sum over $\sigma$ is over all permutations of the external gluons,
the sum over $\pi$ is over all permutations of the colour indices of the antiquarks.
The sum over $\{i_1,...,i_{n_q}\}$ is over all partitions of $n_g$ into $n_q$ non-negative integers and corresponds to the different possibilities to
distribute $n_g$ gluons among $n_q$ open strings. For $n_q > 2$ the partial amplitudes $A^{(0)}_n$ are in general not cyclic ordered.
This is related to the fact that for $n_q \ge 2$ there can be so-called $U(1)$-gluons, corresponding to the second term of the Fierz identity in eq.~(\ref{fierz_identity}).
For $n_q \in \{0,1\}$ the $U(1)$-gluons drop out. In a Born amplitude with $n_q$ quark-antiquark pairs there can be up to $(n_q-1)$ gluons of type $U(1)$.
For the special case $n_q=1$ the colour decomposition of eq.~(\ref{tree_multi_quark}) 
reduces to the colour decomposition of eq.~(\ref{colour_decomp_qqbar_gluon}).

Let us now consider the colour decomposition of one-loop amplitudes.
In the colour decomposition of one-loop amplitudes we can have one additional closed string in comparison to the corresponding
Born amplitude.
Thus the colour decomposition of the one-loop all-gluon amplitude into
partial amplitudes reads \cite{Bern:1990ux}:
\bq
\label{one_loop_all_gluon}
{\cal A}^{(1)}_{n}(g_1,...,g_n) 
 & = & 
 \left(\frac{g}{\sqrt{2}}\right)^n 
 \;\;
 \sum\limits_{m=0}^{\lfloor \frac{n}{2} \rfloor} 
 \;\;
 \sum\limits_{\sigma \in S_{n}/(Z_{n-m} \times Z_{m})} 
 c_{\mathrm{closed}}\left(g_{\sigma_1},...,g_{\sigma_{n-m}}\right)
 c_{\mathrm{closed}}\left(g_{\sigma_{n-m+1}},...,g_{\sigma_n}\right)
 \nonumber \\
 & &
 A^{(1)}_{n,m}\left( g_{\sigma_1}, ..., g_{\sigma_{n-m}}; g_{\sigma_{n-m+1}}, ..., g_{\sigma_n} \right),
\eq
where $\lfloor \frac{n}{2} \rfloor$ denotes the largest integer smaller or equal to $\frac{n}{2}$.
The colour decomposition of an one-loop amplitude containing one or more quark-antiquark pairs is
\bq
\label{one_loop_multi_quark}
\lefteqn{
 {\cal A}^{(1)}_n = 
 \left(\frac{g}{\sqrt{2}}\right)^{n}
} & &
 \nonumber \\
 & &
 \sum\limits_{\pi \in S_{n_q}}
 \sum\limits_{\stackrel{i_1,...,i_{n_q},m \ge 0}{i_1+...+i_{n_q}+m=n_g}}
 \sum\limits_{\sigma \in S_{n_g}/Z_m} 
 c_{\mathrm{open}}\left(q_1,g_{\sigma_1},...,g_{\sigma_{i_1}},\bar{q}_{\pi_1}\right)
 c_{\mathrm{open}}\left(q_2,g_{\sigma_{i_1+1}},...,g_{\sigma_{i_1+i_2}},\bar{q}_{\pi_2}\right)
 \nonumber \\
 & & 
 ...
 c_{\mathrm{open}}\left(q_{n_q},g_{\sigma_{i_1+...+i_{n_q-1}+1}},...,g_{\sigma_{i_1+...+i_{n_q}}},\bar{q}_{\pi_{n_q}}\right)
 c_{\mathrm{closed}}\left(g_{\sigma_{i_1+...+i_{n_q}+1}},...,g_{\sigma_{n_g}}\right)
 \nonumber \\
 & &
 A^{(1)}_{n,m}\left(q_1,g_{\sigma_1},...,g_{\sigma_{i_1}},\bar{q}_{\pi_1}, q_2, ..., g_{\sigma_{i_1+...+i_{n_q}}}, \bar{q}_{\pi_{n_q}}; g_{\sigma_{n_g-m+1}},...,g_{\sigma_{n_g}} \right).
\eq
The partial amplitudes $A^{(1)}_{n,m}$ are in general not cyclic ordered.
This is either due to the additional closed colour string or in the case of amplitudes involving a quark-antiquark pair
due to $U(1)$-gluons.
In a one-loop amplitude with $n_q$ quark-antiquark pairs there can be up to $n_q$ gluons of type $U(1)$.

\subsection{Primitive amplitudes}
\label{subsect:primitive_amplitudes}

The partial amplitudes of the previous sub-section may be further decomposed into smaller objects, called primitive amplitudes.

Tree-level primitive amplitudes are purely kinematic objects, which are gauge-invariant and which have a fixed cyclic ordering of the external
legs.
We will denote tree-level primitive amplitudes by $P^{(0)}$.
They are calculated from planar diagrams with the colour-ordered Feynman rules given in appendix \ref{appendix:colour_ordered_rules}.
These Feynman rules correspond to the colour-stripped Feynman rules of a $U(N)$-gauge theory with quarks in the adjoint representation.
In a $U(N)$-gauge theory additional $U(1)$-gluons are absent, since the Fierz identity for $U(N)$ simply reads
\bq
 T^a_{ij} T^a_{kl} & = &  \frac{1}{2} \delta_{il} \delta_{jk},
 \;\;\;\;\;\;
 \mbox{for $U(N)$.}
\eq
With quarks in the adjoint representation, all colour-ordered three-valent vertices are anti-sym\-metric under the exchange of two of the three external particles.

At the one-loop level primitive one-loop amplitudes $P^{(1)}$ are further specified by two additional properties:
The first property is the particle content inside the loop. This could either be a closed fermion loop or a loop containing at least one gluon or ghost propagator.
In the former case primitive amplitudes with a closed fermion loop are notated by a superscript $P^{(1) [1/2]}$, while in the latter case primitive
amplitudes with at least one gluon or ghost propagator are notated by a superscript $P^{(1) [1]}$.

If the external legs of a one-loop amplitude involve a quark-antiquark pair
we can distinguish the two cases where the loop lies to right or to the left of the fermion line
if we follow the fermion line in the direction of the flow of the fermion number.
We call a fermion line ``left-moving'' if, following the
arrow of the fermion line, the loop is to the right.
Analogously, we call a fermion line ``right-moving'' if, following the
arrow of the fermion line, the loop is to the left.
It turns out that in the decomposition into primitive amplitudes a specific quark line 
is, in all diagrams which contribute to a specific primitive amplitude, either always
left-moving or always right-moving \cite{Bern:1994fz}.
Therefore in the presence of external fermions primitive amplitudes are in addition characterised
by the routing of the fermion lines through the amplitude.

As in the tree-level case the primitive one-loop amplitudes have the properties that they are gauge-invariant and cyclic ordered.

Let us now consider a specific process with $n$ external particles and specified particle identity of the external particles.
In order to simplify the notation, we denote by
\bq
 \left\{ A_i^{(l)} \right\},
 & &
 l \in \{0,1\}
\eq
the set of all tree-level partial amplitudes (for $l=0$) and the set of all one-loop partial amplitudes (for $l=1$).
Similar, we denote by
\bq
 \left\{ P_j^{(l)} \right\},
 & &
 l \in \{0,1\}
\eq
the set of all tree-level primitive amplitudes (for $l=0$) and the set of all one-loop primitive amplitudes (for $l=1$).
The partial amplitudes can be expressed as a linear combination of the primitive amplitudes
\bq
\label{partial_to_primitive}
 A_i^{(l)}
 & = & 
 \sum\limits_j F^{(l)}_{ij} P_j^{(l)}.
\eq
In this paper we present an algorithm based on shuffle relations to determine the coefficients $F^{(l)}_{ij}$ for one-loop (and tree-level) amplitudes.

The relation between partial amplitudes and primitive amplitudes is trivial in the case of the tree-level amplitudes with all gluons or in the case
of the tree-level amplitudes with one quark-antiquark pair and $(n-2)$ gluons.
In these two cases the partial amplitudes are also primitive:
\bq
\label{example_partial_to_primitive}
 A_n^{(0)}\left(g_1,...,g_n\right) & = & P_n^{(0)}\left(g_1 ... g_n\right),
 \nonumber \\
 A_n^{(0)}\left(q,g_1,...,g_{n-2},\bar{q}\right) & = & P_n^{(0)}\left(q g_1 ... g_{n-2} \bar{q}\right).
\eq
We stress that in the general case partial amplitudes are a linear combination of primitive amplitudes.

There is one further complication in the relation of partial amplitudes to primitive amplitudes: 
In general, the linear combination, which expresses a partial amplitude in terms of primitive amplitudes as in eq.~(\ref{partial_to_primitive}) is not
unique.
This can already be seen in the case of tree-level amplitudes with only gluons.
Let
\bq
 w_1 = \alpha_1 \alpha_2 ... \alpha_j,
 & & 
 w_2 = \beta_1 \beta_2 ... \beta_{n-2-j}
\eq
be two ordered sequences of numbers, such that
\bq
 \{1\} \cup \{ \alpha_1, ..., \alpha_j  \} \cup \{ \beta_1, ..., \beta_{n-2-j} \} \cup \{ n \}
 & = &
 \{ 1,...,n \}.
\eq
We further set $w_2^T = \beta_{n-2-j} ... \beta_2 \beta_1$.
\begin{figure}
\begin{center}
\begin{picture}(200,100)(0,0)
 \Line(10,50)(190,50)
 \Line(30,50)(30,90)
 \Line(50,50)(50,10)
 \Line(80,70)(70,90)
 \Line(80,70)(90,90)
 \Line(80,50)(80,70)
 \Line(120,30)(110,10)
 \Line(120,30)(130,10)
 \Line(120,50)(120,30)
 \Line(150,50)(150,90)
 \Line(170,50)(170,10)
 \Text(5,50)[r]{$1$}
 \Text(195,50)[l]{$n$}
 \Text(30,95)[b]{$\alpha_1$}
 \Text(70,95)[b]{$\alpha_2$}
 \Text(90,95)[b]{$\alpha_3$}
 \Text(150,95)[b]{$\alpha_4$}
 \Text(170,5)[t]{$\beta_1$}
 \Text(130,5)[t]{$\beta_2$}
 \Text(110,5)[t]{$\beta_3$}
 \Text(50,5)[t]{$\beta_4$}
\end{picture} 
\caption{\label{figure_Kleiss_Kuijf}
Illustration of the Kleiss-Kuijf relation: 
Only currents consisting of particles from $\alpha_1$, ..., $\alpha_j$
or $\beta_1$, ..., $\beta_{n-2-j}$ couple to the line from $1$ to $n$.
No mixed currents couple to the line from $1$ to $n$.
}
\end{center}
\end{figure}
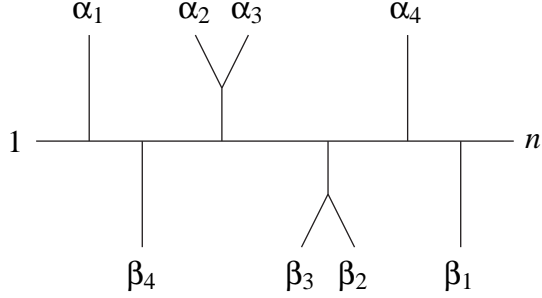
Then
\bq
\label{Kleiss_Kuijf}
 P_n^{(0)}\left( g_1 g_{\alpha_1} ... g_{\alpha_j} g_n g_{\beta_1} ... g_{\beta_{n-2-j}} \right)
 & = & 
 \left( -1 \right)^{n-2-j}
 \sum\limits_{\sigma \in w_1 \; \Sha \; w_2^T}
 P_n^{(0)}\left( g_1 g_{\sigma_1} ... g_{\sigma_{n-2}} g_n \right).
 \nonumber \\
\eq
Eq.~(\ref{Kleiss_Kuijf}) is an example of a Kleiss-Kuijf relation \cite{Kleiss:1988ne,Berends:1989zn,DelDuca:1999rs}. 
We may therefore always replace the primitive amplitude on the left-hand side by the linear combination on the right-hand side.

It is also instructive to review the proof of the Kleiss-Kuijf relation.
It is sufficient to consider a theory with only three-valent vertices, which are anti-symmetric under the exchange of any two legs.
A Yang-Mills theory can be cast into this form by replacing the four-gluon vertex by a gluon-tensor vertex \cite{Draggiotis:1998gr,Duhr:2006iq}.
Now consider the situation shown in fig.~(\ref{figure_Kleiss_Kuijf}), corresponding to a specific contribution to the left-hand side 
of eq.~(\ref{Kleiss_Kuijf}).
Only currents consisting of particles from $\alpha_1$, ..., $\alpha_j$
or $\beta_1$, ..., $\beta_{n-2-j}$ couple to the line from $1$ to $n$.
No mixed currents couple to the line from $1$ to $n$.
Next consider the right-hand side of eq.~(\ref{Kleiss_Kuijf}). Now all particles from 
$\{ \alpha_1, ..., \alpha_j  \} \cup \{ \beta_1, ..., \beta_{n-2-j} \}$ are above the line from $1$ to $n$.
Flipping the particles from the set $\{ \beta_1, ..., \beta_{n-2-j} \}$ will give the sign $(-1)^{n-2-j}$.
The shuffle product will cancel all contributions from mixed currents coupling to the line from $1$ to $n$.
To see this, consider a mixed current contributing to the right-hand side of eq.~(\ref{Kleiss_Kuijf}). 
Such a current will necessarily contain two sub-currents, one made out entirely of particles from
$\{ \alpha_1, ..., \alpha_j  \}$, the other made out entirely of particles from $\{ \beta_1, ..., \beta_{n-2-j} \}$ and coupled together through 
a three-valent vertex.
The shuffle product ensures that both cyclic orderings at this vertex contribute.
Since the three-valent vertex is anti-symmetric under the exchange of two legs, these contributions cancel.
Note that these arguments are not specific to gluons. The arguments apply to any theory with anti-symmetric three-valent vertices only.
These arguments will be at the core of our method.

Let us now consider the letters $l_1$, $l_2$, ..., $l_n$ from the alphabet $\{q_1,q_2,...,\bar{q}_1,\bar{q}_2,...,g_1,g_2,...\}$.
In the case of loop amplitudes the quarks and antiquarks may carry an additional label $L$ or $R$, indicating left- or right-moving fermions.
Let
\bq
 w & = & \left( l_1 l_2 ... l_n \right)
\eq
be a cyclic word. A cyclic word can be used to denote the cyclic order of a primitive amplitude and 
we use the notation
\bq
 P^{(l)}\left(w\right)
 & = &
 P^{(l)}\left( l_1 l_2 ... l_n \right).
\eq
We define the reversed word $w^T$ by
\bq
 \left( l_1 l_2 ... l_n \right)^T
 & = &
 \left. \left( l_n ... l_2 l_1 \right) \right|_{L \leftrightarrow R}.
\eq
The subscript $L \leftrightarrow R$ indicates that in the case that fermion routing labels are present, all fermion routing labels are exchanged.
With this notation we can state the reflection identity for primitive amplitudes:
\bq
\label{reflection_identity}
 P^{(l)}\left( w \right)
 & = &
 \left( -1 \right)^n
 P^{(l)}\left( w^T \right).
\eq
The proof of the reflection identity is simple: Primitive amplitudes are amplitudes derived from a $U(N)$-gauge theory with quarks in the adjoint
representation.
Replacing the four-gluon vertex by a gluon-tensor vertex one arrives at a theory with only three-valent vertices, which are anti-symmetric under the exchange
of any pair of legs.
$P^{(l)}(w^T)$ denotes a primitive amplitude with the reversed cyclic order (and swapped routing labels). This amplitude is obtained from the original one
by changing in every diagram the cyclic order of every vertex.
In a tree amplitude with $n$ external particles there are $(n-2)$ vertices, while in a one-loop amplitude there are $n$ vertices.
In both cases we obtain a factor $(-1)^n$ from the anti-symmetry of the vertices.

\subsection{Decomposition based on Feynman diagrams and linear equations}

There is a general algorithms to find the coefficients $F^{(l)}_{ij}$ which is based on Feynman diagrams and linear equations \cite{Ellis:2008qc,Ellis:2011cr,Ita:2011ar,Badger:2012pg}.
The idea is the following:
Each partial amplitude can be written as a sum of colour-stripped diagrams, with coefficients which are functions of $N$:
\bq
 A_i^{(l)} & = & \sum\limits_k C_{ik}^{(l)} D_k^{(l)}.
\eq
Similar, each primitive amplitude can be written as a sum of colour-stripped diagrams with coefficients $\pm 1$ or $0$:
\bq
 P_j^{(l)} & = & \sum\limits_k M_{jk}^{(l)} D_k^{(l)}.
\eq
Eq.~(\ref{partial_to_primitive}) then becomes
\bq
 \sum\limits_k C_{ik}^{(l)} D_k^{(l)}
 & = & 
 \sum\limits_{j,k} F^{(l)}_{ij} M_{jk}^{(l)} D_k^{(l)},
\eq
and the problem reduces to find a solution of the linear system of equations
\bq
 C_{ik}^{(l)} 
 & = & 
 \sum\limits_{j} F^{(l)}_{ij} M_{jk}^{(l)}.
\eq
This approach provides a method to find the unknown coefficients $F^{(l)}_{ij}$.
However, it is unsatisfactory for the following two reasons:
First of all one has to solve a (large) system of linear equations. A method which avoids this is certainly preferred.
Secondly, the method relies on Feynman diagrams. This is unaesthetic, as all other parts of a next-to-leading order calculation can be performed
without resorting to Feynman diagrams.
There is no compelling reason why Feynman diagrams should be needed in the decomposition of partial amplitudes into primitive amplitudes.
It only reflects the fact, that up to now we do not know a better method.

\section{Basic operations}
\label{sect:operations}

Our method is based on a few basic operations, which we present in this section.
These operations fall into two broad categories, 
related on the one hand to double-ring structures in one-loop colour-flow diagrams and on the other hand to $U(1)$-gluons.
In the former case the reduction of double-ring structures is well known from the pure gluon one-loop amplitude and we briefly review
this operation.
In the latter case the category of basic operations with $U(1)$-gluons is subdivided further into two cases.
We discuss first the case, where we can treat the $U(1)$-gluon with tree-level methods.
In the second case the $U(1)$-gluon appears as a loop propagator and
we define a ``loop closing'' operation by a $U(1)$-gluon.

\subsection{Reduction of a double-ring structure}

Let us consider two cyclic words $u$ and $v$, where
\bq
 u = \left( l_1 ... l_k \right),
 & &
 v = \left( l_{k+1} ... l_r \right),
\eq
and a partial amplitude, where the external legs corresponding to $u$ are cyclic ordered among themselves,
similar the external legs corresponding to $v$ are cyclic ordered among themselves, but no order is implied between legs from $u$ and $v$.
A prominent example are the sub-leading contributions to the one-loop all-gluon amplitude, where $u$ and $v$ correspond to the two 
closed colour strings in eq.~(\ref{one_loop_all_gluon}).
Diagrams contributing to the one-loop all-gluon amplitude can be drawn in the colour-flow representation as a double ring structure, where
for example emission from the outer ring follows the (clockwise) cyclic order of $u$, 
while emission from the inner ring follows the (anti-clockwise) cyclic order of $v$.
We would like to express such amplitudes as a linear combination of amplitudes with only one overall cyclic order.
This can be done with the help of the cyclic shuffle product:
The expression
\bq
\label{resolve_double_ring}
 \left(-1\right)^{r-k} \; u \circledcirc v^T
\eq
gives the correct linear combination of cyclic orderings.
The proof of eq.~(\ref{resolve_double_ring}) is just the cyclic version of the proof of the Kleiss-Kuijf relation,
presented at the end of section~\ref{subsect:primitive_amplitudes}.

\subsection{The exchange of a $U(1)$-gluon as a tree-like object}

We now consider $U(1)$-gluons. We start our discussion with the case, where a $U(1)$-gluon is exchanged as a tree-like object.
This applies first of all to Born amplitudes with two or more quark-antiquark pairs.
This case is discussed in~\ref{subsect:U1_tree_exchange_1}.
There are two slight generalisations of this case:
First we can consider the case, where one quark line is colour-connected to a loop.
This case is discussed in~\ref{subsect:U1_tree_exchange_2}. The $U(1)$-gluon remains a tree-like object.
Secondly, we have the case, where a $U(1)$-gluon is exchanged between a quark line and a closed quark loop.
This case is discussed in~\ref{subsect:U1_tree_exchange_3}.
Again, the $U(1)$-gluon does not enter the loop and remains a tree-like object.

\subsubsection{The exchange of a $U(1)$-gluon between two tree-like quark lines}
\label{subsect:U1_tree_exchange_1}

Let us consider two cyclic words $u$ and $v$, which both contain a quark-antiquark pair.
We write
\bq
 u = \left( l_1 ... l_k \right) = \left( \bar{q}_i u_{i,L} q_i u_{i,R} \right),
 & &
 v = \left( l_{k+1} ... l_r \right) = \left( \bar{q}_j v_{j,L} q_j v_{j,R} \right),
\eq
where $u_{i,L}$, $u_{i,R}$, $v_{j,L}$ and $v_{j,R}$ are words from the alphabet
$A=\{\bar{q}_1,\bar{q}_2,...,q_1,q_2,...,g_1,g_2,...\}$ not containing the letters $\bar{q}_i$, $q_i$, $\bar{q}_j$ and $q_j$.
We can think of the cyclic words $u$ and $v$ as describing two tree-like colour clusters. 
We now define a operation $U_{ij}(u,v)$,
which corresponds to the decomposition of an amplitude with an 
exchange of a $U(1)$-gluon between the quark line $i$ and the quark line $j$ 
into primitive parts.
The definition of $U_{ij}(u,v)$ is not unique, reflecting the fact that the decomposition of partial amplitudes into primitive amplitudes
is not unique.
One possibility is to define
\bq
\label{def_U1_shuffle_ij}
 U_{ij}\left( u, v \right)
 & = &
 \sum\limits_{\stackrel{(\mbox{\tiny cyclic shuffles} \; \sigma) / {\mathbb Z}_r}{(q_i ... \bar{q}_i ... q_j ... \bar{q}_j ...)}} \left( l_{\sigma(1)} l_{\sigma(2)} ... l_{\sigma(r)} \right),
\eq
where the sum is over all cyclic shuffles with the cyclic order of the quarks/antiquarks given by $(q_i ... \bar{q}_i ... q_j ... \bar{q}_j ... )$.
This means that the other possible cyclic orders 
\bq
(q_i ... \bar{q}_i ... \bar{q}_j ... q_j ...),
 \;\;\;
(\bar{q}_i ... q_i ... q_j ... \bar{q}_j ...)
 & \mbox{and} &
(\bar{q}_i ... q_i ... \bar{q}_j ... q_j ...)
\eq
are excluded.
In appendix~\ref{appendix:U1_shuffle} we prove
that $U_{ij}$ corresponds to the exchange of a $U(1)$-gluon between the two quark lines $i$ and $j$
(this gives the reason why we denote this operation with the letter ``$U$'').
Eq.~(\ref{def_U1_shuffle_ij}) gives the decomposition into primitive amplitudes for this case.
As already noted above, this decomposition is in general not unique.
We could have defined $U_{ij}$ equally well by
\bq
\label{other_possibilities}
 & & 
 - \sum\limits_{\stackrel{(\mbox{\tiny cyclic shuffles} \; \sigma) / {\mathbb Z}_r}{(q_i ... \bar{q}_i ... \bar{q}_j ... q_j ...)}} 
  \left( l_{\sigma(1)} l_{\sigma(2)} ... l_{\sigma(r)} \right),
 \nonumber \\
 & & 
 - \sum\limits_{\stackrel{(\mbox{\tiny cyclic shuffles} \; \sigma) / {\mathbb Z}_r}{(\bar{q}_i ... q_i ... q_j ... \bar{q}_j ... )}} 
  \left( l_{\sigma(1)} l_{\sigma(2)} ... l_{\sigma(r)} \right)
 \;\;\;\;\;\;
 \mbox{or}
 \nonumber \\
 & & 
 \sum\limits_{\stackrel{(\mbox{\tiny cyclic shuffles} \; \sigma) / {\mathbb Z}_r}{(\bar{q}_i ... q_i ... \bar{q}_j ... q_j ...)}} 
  \left( l_{\sigma(1)} l_{\sigma(2)} ... l_{\sigma(r)} \right).
\eq
Each possibility gives a different linear combination of primitive amplitudes, which however reduces to the same set of diagrams.
Eq.~(\ref{def_U1_shuffle_ij}) corresponds to the case, where the $U(1)$-gluon is emitted from both quark lines to the left, while
the three other possibilities in eq.~(\ref{other_possibilities}) corresponds to the cases where the $U(1)$-gluon is emitted from one or both
quark lines to the right.
The choice in eq.~(\ref{def_U1_shuffle_ij}) generates the minimal number of primitive amplitudes if in the initial cyclic words $u$ and $v$
all other particles are emitted to the right of the quark lines.

There is even more freedom in defining $U_{ij}(u,v)$. We could have equally well have defined
\bq
\label{def_U1tilde_shuffle_ij}
 \tilde{U}_{ij}\left(u,v\right)
 & = &
 \left( -1 \right)^{n_{u_{i,L}}+n_{v_{j,L}}}
 \sum\limits_{w \in \left( q_i \left( u_{i,R} \Sha u_{i,L}^T \right) \bar{q}_i q_j \left( v_{j,R} \Sha v_{j,L}^T \right) \bar{q}_j \right)}
 w,
\eq
where $n_{u_{i,L}}$ is the length of the word $u_{i,L}$ and $n_{v_{j,L}}$ is the length of the word $v_{j,L}$.
The proof of eq.~(\ref{def_U1tilde_shuffle_ij}) is rather short: Eq.~(\ref{def_U1tilde_shuffle_ij}) is trivial if both
$u_{i,L}$ and $v_{j,L}$ are empty.
In the case where one of them or both are non-empty, one recognises in the expressions
\bq
 \left( -1 \right)^{n_{u_{i,L}}} \left( u_{i,R} \Sha u_{i,L}^T \right)
 & \mbox{and} &
 \left( -1 \right)^{n_{v_{j,L}}} \left( v_{j,R} \Sha v_{j,L}^T \right)
\eq
the Kleiss-Kuijf relation, which flips the legs of $u_{i,L}$ and $v_{j,L}$ to the other side and reduces the problem to the trivial case.
The difference between $U_{ij}$ and $\tilde{U}_{ij}$ lies in the fact that for $U_{ij}$ the legs of $u_{i,L}$ remain on the same side of the
quark line $i$, while for $\tilde{U}_{ij}$ they are flipped to the other side of the quark line $i$.
Similar, for $U_{ij}$ the legs of $v_{j,L}$ remain on the same side of the quark line $j$, while for $\tilde{U}_{ij}$ they are flipped to the other
side of the quark line $j$.
In analogy with eq.~(\ref{other_possibilities}) there are also for $\tilde{U}_{ij}$ three other possible definitions obtained by exchanging left and right
for the quark line $i$, the quark line $j$ or both.

The four possibilities of defining $U_{ij}$ and the four possibilities of defining $\tilde{U}_{ij}$ do not exhaust our freedom
in the decomposition of a partial amplitude with a $U(1)$-gluon into primitive amplitudes.
We can also treat one quark line as in eq.~(\ref{def_U1_shuffle_ij}), while the other is treated as in eq.~(\ref{def_U1tilde_shuffle_ij}).
This freedom will be useful in loop amplitudes.

Now suppose that $u$ contains $n_{q_u}$ quark-antiquark pairs labelled by $ i \in \{1, 2, ..., n_{q_u}\}$ 
and that $v$ contains $n_{q_v}$ quark-antiquark pairs labelled by $j \in \{n_{q_u}+1, n_{q_u}+2, ..., n_{q_u}+n_{q_v}\}$.
We then set
\bq
\label{def_U1_shuffle}
 U\left( u, v \right)
 & = &
 \sum\limits_{i \in K}
 \;\;
 \sum\limits_{j \in L}
 \;
 U_{ij}\left( u, v \right),
\eq
with $K=\{1, 2, ..., n_{q_u}\}$ and $L=\{n_{q_u}+1, n_{q_u}+2, ..., n_{q_u}+n_{q_v}\}$.
The definition of $U(u_1,u_2,...,u_k)$ with more than two arguments is more involved.
Suppose that $u_i$ contains $n_{q_i}$ quark-antiquark pairs labelled from $n_{q_1}+...+n_{q_{i-1}}+1$ to $n_{q_1}+...+n_{q_{i-1}}+n_{q_i}$.
The operation $U(u_1,u_2,...,u_k)$ is defined by the following algorithm:
\\
\\
Algorithm 1:
\begin{enumerate}
\item Draw all connected tree diagrams with $k$ vertices, labelled by $u_1$, $u_2$, ..., $u_k$.
\item For every edge connecting $u_k$ and $u_l$ use an operation $U_{ij}$, where $i$ labels a quark-antiquark pair in $u_k$ and
$j$ labels a quark-antiquark pair in $u_l$ and sum over $i$ and $j$.
The order in which these operations are performed is irrelevant, since the operations for a given graph are associative.
\item Sum over all diagrams.
\end{enumerate}
The reason for this on the first sight rather complicated definition is the following: A naive iteration of the operation with two factors 
in eq.~(\ref{def_U1_shuffle}) is not associative. To see this we take as an example the three cyclic words
\bq
 u_1 = \left( \bar{q}_1 q_1 \right),
 \;\;\;\;\;\;
 u_2 = \left( \bar{q}_2 q_2 \right),
 \;\;\;\;\;\;
 u_3 = \left( \bar{q}_3 q_3 \right)
\eq
and consider $U( U(u_1,u_2), u_3)$ and $U( u_1, U(u_2,u_3))$. 
We have
\bq
\lefteqn{
 U\left( U\left(u_1,u_2\right), u_3 \right)
 = }
 & & \nonumber \\
 & &
 \left( q_1 \bar{q}_1 q_2 \bar{q}_2 q_3 \bar{q}_3 \right)
 +
 \left( q_1 \bar{q}_1 q_3 \bar{q}_3 q_2 \bar{q}_2 \right)
 +
 \left( q_1 \bar{q}_1 q_2 q_3 \bar{q}_3 \bar{q}_2 \right)
 +
 \left( q_1 \bar{q}_1 q_3 q_2 \bar{q}_2 \bar{q}_3 \right)
 \nonumber \\
 & &
 +
 \left( q_2 \bar{q}_2 q_1 \bar{q}_1 q_3 \bar{q}_3 \right)
 +
 \left( q_2 \bar{q}_2 q_3 \bar{q}_3 q_1 \bar{q}_1 \right)
 +
 \left( q_2 \bar{q}_2 q_1 q_3 \bar{q}_3 \bar{q}_1 \right)
 +
 \left( q_2 \bar{q}_2 q_3 q_1 \bar{q}_1 \bar{q}_3 \right),
 \nonumber \\
\lefteqn{
 U\left( u_1, U\left( u_2, u_3 \right) \right)
 = }
 & & \nonumber \\
 & &
 \left( q_2 \bar{q}_2 q_1 \bar{q}_1 q_3 \bar{q}_3 \right)
 +
 \left( q_2 \bar{q}_2 q_3 \bar{q}_3 q_1 \bar{q}_1 \right)
 +
 \left( q_2 \bar{q}_2 q_1 q_3 \bar{q}_3 \bar{q}_1 \right)
 +
 \left( q_2 \bar{q}_2 q_3 q_1 \bar{q}_1 \bar{q}_3 \right)
 \nonumber \\
 & &
 +
 \left( q_3 \bar{q}_3 q_2 \bar{q}_2 q_1 \bar{q}_1 \right)
 +
 \left( q_3 \bar{q}_3 q_1 \bar{q}_1 q_2 \bar{q}_2 \right)
 +
 \left( q_3 \bar{q}_3 q_2 q_1 \bar{q}_1 \bar{q}_2 \right)
 +
 \left( q_3 \bar{q}_3 q_1 q_2 \bar{q}_2 \bar{q}_1 \right).
\eq
$U( U(u_1,u_2), u_3)$ reduces to diagrams, where a $U(1)$-gluon connects the quark lines $1$ and $2$, 
while a second $U(1)$-gluon connects the quark line $3$ either with quark line $1$ or $2$.
On the other hand, $U( u_1, U(u_2,u_3))$ reduces to diagrams, 
where a $U(1)$-gluon connects the quark lines $2$ and $3$, 
while a second $U(1)$-gluon connects the quark line $1$ either with quark line $2$ or $3$.
We would like to have that the correct result reduces to diagrams, which connect the three quark lines in all possible ways by two $U(1)$-gluons.
The above algorithm gives the correct result:
\bq
\label{long_result}
\lefteqn{
 U\left( u_1, u_2, u_3 \right)
 = } & & \nonumber \\
 & &
 \left( q_1 \bar{q}_1 q_2 \bar{q}_2 q_3 \bar{q}_3 \right)
 +
 \left( q_1 \bar{q}_1 q_3 \bar{q}_3 q_2 \bar{q}_2 \right)
 +
 \left( q_1 \bar{q}_1 q_2 q_3 \bar{q}_3 \bar{q}_2 \right)
 +
 \left( q_1 \bar{q}_1 q_3 q_2 \bar{q}_2 \bar{q}_3 \right)
 \nonumber \\
 & &
 +
 \left( q_2 \bar{q}_2 q_1 \bar{q}_1 q_3 \bar{q}_3 \right)
 +
 \left( q_2 \bar{q}_2 q_3 \bar{q}_3 q_1 \bar{q}_1 \right)
 +
 \left( q_2 \bar{q}_2 q_1 q_3 \bar{q}_3 \bar{q}_1 \right)
 +
 \left( q_2 \bar{q}_2 q_3 q_1 \bar{q}_1 \bar{q}_3 \right),
 \nonumber \\
 & &
 +
 \left( q_3 \bar{q}_3 q_2 \bar{q}_2 q_1 \bar{q}_1 \right)
 +
 \left( q_3 \bar{q}_3 q_1 \bar{q}_1 q_2 \bar{q}_2 \right)
 +
 \left( q_3 \bar{q}_3 q_2 q_1 \bar{q}_1 \bar{q}_2 \right)
 +
 \left( q_3 \bar{q}_3 q_1 q_2 \bar{q}_2 \bar{q}_1 \right).
\eq
For this reason we do not denote the operation $U(u_1,u_2,...,u_k)$ with multiplication operators, but use a notation in the form of a function with $k$ arguments instead.
We also note that shuffle relations do not necessarily generate the most compact result. This can be seen in the example of eq.~(\ref{long_result}):
In this example the sum of the last ten terms on the right-hand side of eq.~(\ref{long_result}) will be a complicated zero, when the cyclic words are taken
as the arguments of primitive tree amplitudes.

\subsubsection{The exchange of a $U(1)$-gluon between a tree-like quark line and a quark line colour-connected to a loop}
\label{subsect:U1_tree_exchange_2}

Next we consider the case of the exchange of a $U(1)$-gluon between two external quark lines $i$ and $j$, where the quark line
$i$ is colour-connected to a loop.
The loop can either be a closed fermion, in which case we use a superscript $[1/2]$, or a loop containing at least one gluon
or ghost propagator, in which case we use a superscript $[1]$.
We let
\bq
 u^{[s]} = \left( l_1 ... l_k \right) = \left( \bar{q}_i^{L/R} u_{i,L} q_i^{L/R} u_{i,R} \right),
 & &
 v = \left( l_{k+1} ... l_r \right) = \left( \bar{q}_j v_{j,L} q_j v_{j,R} \right)
\eq
and we assume that the primitive amplitude corresponding to $u^{[s]}$ is colour-connected to the loop.
The superscript $[s] \in \{ [1/2], [1] \}$ indicates the particle content in the loop.
Quarks and antiquarks in $u^{[s]}$ have routing labels $L$ or $R$ assigned, while for quarks and antiquarks in $v$ the routing labels
are not yet specified.
We distinguish the two cases, whether the quark line $i$ is left- or right-moving.
We start with the $q_i^L$ case. 
In this case the loop is to the right of the quark line $i$ and we would like to leave the cyclic order $...q_i^L u_{i,R} \bar{q}_i^L ...$
fixed.
Therefore we treat the quark line $i$ as in eq.~(\ref{def_U1_shuffle_ij}), while the quark line $j$ is treated as in eq.~(\ref{def_U1tilde_shuffle_ij}).
We arrive at
\bq
\label{def_ij_loop_1}
 U_{ij}\left(u^{[s]},v\right)
 =  
 U_{ji}\left(v,u^{[s]}\right)
 & = &
 \left(-1\right)^{n_{v_{j,L}}}
 \sum\limits_{w \in \left( q_i^L u_{i,R} \bar{q}_i^L \left( \left( q_j^R \left( v_{j,R} \Sha v_{j,L}^T \right) \bar{q}_j^R \right) \Sha u_{i,L}\right)\right)} 
 w.
\eq
The number $n_{v_{j,L}}$ denotes the length of the word $v_{j,L}$.
With this definition the quark line $j$ is necessarily right-moving. 
Additional quark lines, which might be present in $v_{j,R}$ or $v_{j,L}$, have their routing labels assigned as follows:
A quark line $k$, which appears in the cyclic order $q_j^R ... q_k ... \bar{q}_k \bar{q}_j^R$ in $w$ is right-moving, 
while a quark line $k$, which appears in the cyclic order $q_j^R ... \bar{q}_k ... q_k \bar{q}_j^R$ in $w$ is left-moving.
The definition in eq.~(\ref{def_ij_loop_1}) is not unique. We could have equally well used a definition, where the quark line $j$ is always left-moving.

We then consider the $q_i^R$ case.
In this case the loop is to the left of the quark line $i$ and we would like to leave the cyclic order $...\bar{q}_i^R u_{i,L} q_i^R ...$
fixed.
We let
\bq
\label{def_ij_loop_2}
 U_{ij}\left(u^{[s]},v\right)
 =  
 U_{ji}\left(v,u^{[s]}\right)
 & = &
 \left(-1\right)^{n_{v_{j,R}}}
 \sum\limits_{w \in \left( \bar{q}_i^R u_{i,L} q_i^R \left( \left( \bar{q}_j^L \left( v_{j,L} \Sha v_{j,R}^T \right) q_j^L \right) \Sha u_{i,R}\right)\right)} 
 w.
\eq
The number $n_{v_{j,R}}$ denotes the length of the word $v_{j,R}$.
With this definition the quark line $j$ is necessarily left-moving. 
Additional quark lines, which might be present in $v_{j,R}$ or $v_{j,L}$, have their routing labels assigned as follows:
A quark line $k$, which appears in the cyclic order $\bar{q}_j^L ... q_k ... \bar{q}_k q_j^L$ in $w$ is right-moving, 
while a quark line $k$, which appears in the cyclic order $\bar{q}_j^L ... \bar{q}_k ... q_k q_j^L$ in $w$ is left-moving.
Again, the definition in eq.~(\ref{def_ij_loop_2}) is not unique. We could have equally well used a definition, where the quark line $j$ is always right-moving.

Algorithm 1 has a straightforward extension towards quark lines colour-connected to a loop.
One simply uses the definition in eq.~(\ref{def_ij_loop_1}) or eq.~(\ref{def_ij_loop_2}) where appropriate.

\subsubsection{The exchange of a $U(1)$-gluon between a tree-like quark line and a closed quark loop}
\label{subsect:U1_tree_exchange_3}

A $U(1)$-gluon can also be exchanged between an external quark line $i$ and a closed fermion loop.
We let
\bq
 u = \left( l_1 ... l_k \right) = \left( \bar{q}_i u_{i,L} q_i u_{i,R} \right),
 & &
 v^{[1/2]} = \left( l_{k+1} ... l_r \right),
\eq
where $u$ contains the external quark line $i$ and $v$ corresponds to a primitive amplitude with a closed fermion loop.
We assume that the fermion line of the closed fermion loop is counter-clockwise.
For $v^{[1/2]}$ we allow the possibility that the cyclic word is empty: $v^{[1/2]}=()$.
External quarks and antiquarks in $v^{[1/2]}$ have routing labels $L$ or $R$ assigned, while for quarks and antiquarks in $u$ the routing labels
are not yet specified.
We set
\bq
\label{U1_shuffle_i_loop}
 U_{i,\mathrm{loop}}\left(u,v^{[1/2]}\right)
 = 
 U_{\mathrm{loop},i}\left(v^{[1/2]},u\right)
 & = &
 \left(-1\right)^{n_{u_{i,R}}}
 \sum\limits_{\left( \bar{q}_i^L \left( u_{i,L} \Sha u_{i,R}^T \right) q_i^L \right) \circledcirc v}
 w.
\eq
The number $n_{u_{i,R}}$ denotes the length of the word $u_{i,R}$.
With this definition the quark line $i$ is necessarily left-moving. 
Additional quark lines, which might be present in $u_{i,L}$ or $u_{i,R}$, have their routing labels assigned as follows:
A quark line $k$, which appears in the cyclic order $\bar{q}_i^L ... q_k ... \bar{q}_k q_i^L$ in $w$ is right-moving, 
while a quark line $k$, which appears in the cyclic order $\bar{q}_i^L ... \bar{q}_k ... q_k q_i^L$ in $w$ is left-moving.
The definition in eq.~(\ref{U1_shuffle_i_loop}) is not unique. We could have equally well used a definition, where the quark line $i$ is always right-moving.

Algorithm 1 has a straightforward extension towards one-loop amplitudes with a closed fermion loop: One simply treats the closed fermion loop
as an additional fermion line and uses the above defined extension $U_{i,\mathrm{loop}}$ as appropriate.

\subsection{The loop closing operation with a $U(1)$-gluon}

Up to now we considered $U(1)$-gluons only in tree amplitudes or in tree parts of loop amplitudes.
We now turn our attention to $U(1)$-gluons appearing as loop propagators.
We first consider in section~\ref{subsect:loop_closing_1} the case, where a $U(1)$-gluon is emitted and absorbed from the same quark line.
We then discuss in section~\ref{subsect:loop_closing_2} the case, where a $U(1)$-gluon closes a loop between two quark lines.
Symmetry factors for more than one $U(1)$-gluon in the loop are discussed in section~\ref{subsect:loop_closing_3}.

\subsubsection{The loop closing operation on a single quark line}
\label{subsect:loop_closing_1}

We start with the simplest case of the loop closing operation: A $U(1)$-gluon is emitted and absorbed from the same quark line $i$.
We consider a cyclic word 
\bq
 u & = & 
 \left( q_i u_1 \bar{q}_i u_2 \right)
 =
 \left( q_i l_1 ... l_k \bar{q}_i l_{k+1} ... l_r \right),
\eq
with subwords $u_1=l_1...l_k$ and $u_2=l_{k+1}...l_r$ of length $k$ and $(r-k)$, respectively.
Routing labels are not yet assigned.
We set
\bq
\label{U1_closing_ii}
 C_{ii}\left(u\right)
 & = &
 \left(-1\right)^{n_{u_2}}
 \sum\limits_{w \in \left( q_i^R \left( u_1 \Sha u_2^T \right) \bar{q}_i^R \right)}
 w,
\eq
where $n_{u_2}=r-k$ is the length of the word $u_2$.
With this definition, the quark line $i$ is right-moving.
Additional quark lines, which might be present in $u_1$ or $u_2$, have their routing labels assigned as follows:
A quark line $k$, which appears in the cyclic order $q_i^R ... q_k ... \bar{q}_k ... \bar{q}_i^R$ in $w$ is right-moving, 
while a quark line $k$, which appears in the cyclic order $q_i^R ... \bar{q}_k ... q_k ... \bar{q}_i^R$ in $w$ is left-moving.
The definition in eq.~(\ref{U1_closing_ii}) is not unique. We could equally well used a definition, where the quark line $i$ is always left-moving.

In appendix~\ref{appendix:loop_closing} we show that eq.~(\ref{U1_closing_ii}) does indeed correspond to the loop closing operation
by a $U(1)$-gluon on a single quark line.

\subsubsection{The loop closing operation between two quark lines}
\label{subsect:loop_closing_2}

We now consider the loop closing operation through a $U(1)$-gluon between two quark lines $i$ and $j$.
To this aim we first consider the cyclic word
\bq
\label{loop_closing_cyclic_order_1}
 u & = & 
 \left( q_i u_1 \bar{q}_i u_2 q_j u_3 \bar{q}_j u_4 \right),
\eq
with subwords $u_1$, $u_2$, $u_3$ and $u_4$ of length $n_{u_1}$, $n_{u_2}$, $n_{u_3}$ and $n_{u_4}$, respectively.
Routing labels are not yet assigned.
We define the loop closing operation $C_{ij}(u)$ as a sum of four terms:
\bq
\label{def_C_ij}
 C_{ij}\left(u\right)
 & = &
 C_{ij}^{RR}\left(u\right)
 +
 C_{ij}^{RL}\left(u\right)
 +
 C_{ij}^{LR}\left(u\right)
 +
 C_{ij}^{LL}\left(u\right).
\eq
$C_{ij}^{RR}(u)$ is defined by
\bq
\label{def_C_ij_RR}
 C_{ij}^{RR}\left(u\right)
 & = &
 \left(-1\right)^{n_{u_4}}
 \sum\limits_{w \in \left( q_i^R \left(\left( u_1 \bar{q}_i^R u_2 q_j^R u_3 \right) \Sha u_4^T \right) \bar{q}_j^R \right)}
 w.
\eq
In this contribution, the quark lines $i$ and $j$ are right-moving.
Additional quark lines, which might be present in $u_1$, $u_2$, $u_3$ or $u_4$, have their routing labels assigned as follows:
A quark line $k$, which appears in the cyclic order $q_i^R ... q_k ... \bar{q}_k ... \bar{q}_j^R$ in $w$ is right-moving, 
while a quark line $k$, which appears in the cyclic order $q_i^R ... \bar{q}_k ... q_k ... \bar{q}_j^R$ in $w$ is left-moving.

$C_{ij}^{RL}(u)$ is defined by
\bq
\label{def_C_ij_RL}
 C_{ij}^{RL}\left(u\right)
 & = &
 \left(-1\right)^{n_{u_3}+n_{u_4}}
 \sum\limits_{w \in \left( q_i^R \left(\left( u_1 \bar{q}_i^R u_2 \right) \Sha \left( u_4^T \bar{q}_j^L u_3^T \right) \right) q_j^L \right)}
 w.
\eq
In this contribution, the quark line $i$ is right-moving, while the quark line $j$ is left-moving.
Additional quark lines, which might be present in $u_1$, $u_2$, $u_3$ or $u_4$, have their routing labels assigned as follows:
A quark line $k$, which appears in the cyclic order $q_i^R ... q_k ... \bar{q}_k ... q_j^L$ in $w$ is right-moving, 
while a quark line $k$, which appears in the cyclic order $q_i^R ... \bar{q}_k ... q_k ... q_j^L$ in $w$ is left-moving.

$C_{ij}^{LR}(u)$ is defined by
\bq
\label{def_C_ij_LR}
 C_{ij}^{LR}\left(u\right)
 & = &
 \left(-1\right)^{n_{u_1}+n_{u_4}}
 \sum\limits_{w \in \left( \bar{q}_i^L \left(\left( u_2 q_j^R u_3 \right) \Sha \left( u_1^T q_i^L u_4^T \right) \right) \bar{q}_j^R \right)}
 w.
\eq
In this contribution, the quark line $i$ is left-moving, while the quark line $j$ is right-moving.
Additional quark lines, which might be present in $u_1$, $u_2$, $u_3$ or $u_4$, have their routing labels assigned as follows:
A quark line $k$, which appears in the cyclic order $\bar{q}_i^L ... q_k ... \bar{q}_k ... \bar{q}_j^R$ in $w$ is right-moving, 
while a quark line $k$, which appears in the cyclic order $\bar{q}_i^L ... \bar{q}_k ... q_k ... \bar{q}_j^R$ in $w$ is left-moving.

$C_{ij}^{LL}(u)$ is defined by
\bq
\label{def_C_ij_LL}
 C_{ij}^{LL}\left(u\right)
 & = &
 \left(-1\right)^{n_{u_1}+n_{u_3}+n_{u_4}}
 \sum\limits_{w \in \left( \bar{q}_i^L \left( u_2 \Sha \left( u_1^T q_i^L u_4^T \bar{q}_j^L u_3^T \right) \right) q_j^L \right)}
 w.
\eq
In this contribution, the quark lines $i$ and $j$ are left-moving.
Additional quark lines, which might be present in $u_1$, $u_2$, $u_3$ or $u_4$, have their routing labels assigned as follows:
A quark line $k$, which appears in the cyclic order $\bar{q}_i^L ... q_k ... \bar{q}_k ... q_j^L$ in $w$ is right-moving, 
while a quark line $k$, which appears in the cyclic order $\bar{q}_i^L ... \bar{q}_k ... q_k ... q_j^L$ in $w$ is left-moving.

The equations~(\ref{def_C_ij_RR}) to~(\ref{def_C_ij_LL}) give one possibility of defining the operations
$C_{ij}^{RR}$, $C_{ij}^{RL}$, $C_{ij}^{LR}$ and $C_{ij}^{LL}$. These definitions are not unique.
We notice that in eqs.~(\ref{def_C_ij_RR}) to~(\ref{def_C_ij_LL}) the sequence $u_4$ is always reversed, while the sequence
$u_2$ is never reversed.
We could have given an alternative definition, where the sequence $u_2$ is always reversed, while the the sequence $u_4$
is never reversed.

In appendix~\ref{appendix:loop_closing} we show that equations~(\ref{def_C_ij_RR}) to~(\ref{def_C_ij_LL}) do indeed correspond to the loop closing operation
by a $U(1)$-gluon between quark lines $i$ and $j$.

We started the discussion of the loop closing operation between two quark lines with the cyclic order as in eq.~(\ref{loop_closing_cyclic_order_1}).
Beside that one, there are two other possible cyclic orders of the quarks and antiquarks:
\bq
 \left( \bar{q}_i u_1 q_i u_2 \bar{q}_j u_3 q_j u_4 \right),
 & &
 \left( q_i u_1 \bar{q}_i u_2 \bar{q}_j u_3 q_j u_4 \right).
\eq
The case $( \bar{q}_i u_1 q_i u_2 q_j u_3 \bar{q}_j u_4 )$ is cyclic equivalent to $( q_i u_1 \bar{q}_i u_2 \bar{q}_j u_3 q_j u_4 )$.
In the case
\bq
 u & = &
 \left( \bar{q}_i u_1 q_i u_2 \bar{q}_j u_3 q_j u_4 \right)
\eq
we have
\bq
 C_{ij}^{LL}\left(u\right)
 & = &
 \left(-1\right)^{n_{u_4}}
 \sum\limits_{w \in \left( \bar{q}_i^L \left(\left( u_1 q_i^L u_2 \bar{q}_j^L u_3 \right) \Sha u_4^T \right) q_j^L \right)}
 w,
 \nonumber \\
 C_{ij}^{LR}\left(u\right)
 & = &
 \left(-1\right)^{n_{u_3}+n_{u_4}}
 \sum\limits_{w \in \left( \bar{q}_i^L \left(\left( u_1 q_i^L u_2 \right) \Sha \left( u_4^T q_j^R u_3^T \right) \right) \bar{q}_j^R \right)}
 w,
 \nonumber \\
 C_{ij}^{RL}\left(u\right)
 & = &
 \left(-1\right)^{n_{u_1}+n_{u_4}}
 \sum\limits_{w \in \left( q_i^R \left(\left( u_2 \bar{q}_j^L u_3 \right) \Sha \left( u_1^T \bar{q}_i^R u_4^T \right) \right) q_j^L \right)}
 w,
 \nonumber \\
 C_{ij}^{RR}\left(u\right)
 & = &
 \left(-1\right)^{n_{u_1}+n_{u_3}+n_{u_4}}
 \sum\limits_{w \in \left( q_i^R \left( u_2 \Sha \left( u_1^T \bar{q}_i^R u_4^T q_j^R u_3^T \right) \right) \bar{q}_j^R \right)}
 w.
\eq
Additional quark lines, which might be present in $u_1$, $u_2$, $u_3$ or $u_4$, have their routing labels assigned as follows:
A quark line $k$, which appears in the order $... q_k ... \bar{q}_k ...$ is right-moving, 
while a quark line $k$, which appears in the order $... \bar{q}_k ... q_k ... $ is left-moving.

It remains to discuss the case
\bq
 u & = &
  \left( q_i u_1 \bar{q}_i u_2 \bar{q}_j u_3 q_j u_4 \right).
\eq
Here we have
\bq
 C_{ij}^{RL}\left(u\right)
 & = &
 -
 \left(-1\right)^{n_{u_4}}
 \sum\limits_{w \in \left( q_i^R \left(\left( u_1 \bar{q}_i^R u_2 \bar{q}_j^L u_3 \right) \Sha u_4^T \right) q_j^L \right)}
 w,
 \nonumber \\
 C_{ij}^{RR}\left(u\right)
 & = &
 -
 \left(-1\right)^{n_{u_3}+n_{u_4}}
 \sum\limits_{w \in \left( q_i^R \left(\left( u_1 \bar{q}_i^R u_2 \right) \Sha \left( u_4^T q_j^R u_3^T \right) \right) \bar{q}_j^R \right)}
 w,
 \nonumber \\
 C_{ij}^{LL}\left(u\right)
 & = &
 -
 \left(-1\right)^{n_{u_1}+n_{u_4}}
 \sum\limits_{w \in \left( \bar{q}_i^L \left(\left( u_2 \bar{q}_j^L u_3 \right) \Sha \left( u_1^T q_i^L u_4^T \right) \right) q_j^L \right)}
 w,
 \nonumber \\
 C_{ij}^{LR}\left(u\right)
 & = &
 -
 \left(-1\right)^{n_{u_1}+n_{u_3}+n_{u_4}}
 \sum\limits_{w \in \left( \bar{q}_i^L \left( u_2 \Sha \left( u_1^T q_i^L u_4^T q_j^R u_3^T \right) \right) \bar{q}_j^R \right)}
 w.
\eq
Additional quark lines, which might be present in $u_1$, $u_2$, $u_3$ or $u_4$, have their routing labels assigned as follows:
A quark line $k$, which appears in the order $... q_k ... \bar{q}_k ...$ is right-moving, 
while a quark line $k$, which appears in the order $... \bar{q}_k ... q_k ... $ is left-moving.

\subsubsection{Symmetry factors for more than one $U(1)$-gluon in the loop}
\label{subsect:loop_closing_3}

Up to now we considered only the loop closing operation for a single cyclic word.
We also have to consider the case, that more than one $U(1)$-gluon appears as a loop propagator.
Cutting one $U(1)$-gluon gives a tree and we expect that the remaining $U(1)$-gluons can be treated with the operation $U_{ij}$ described
in section~(\ref{subsect:U1_tree_exchange_1}).
This is indeed the case, provided we take symmetry factors into account in order to avoid double counting.
We now define an operation $CU(u_1,...,u_k)$ of $k$ arguments $(k\ge1)$, which combines the operation $U_{ij}$ and $C_{ij}$ (including $C_{ii}$).
We assume that $u_1$, ..., $u_k$ are cyclic words and that
$u_i$ contains $n_{q_i}$ quark-antiquark pairs labelled from $n_{q_1}+...+n_{q_{i-1}}+1$ to $n_{q_1}+...+n_{q_{i-1}}+n_{q_i}$.
The operation $CU(u_1,u_2,...,u_k)$ is defined by the following algorithm:
\\
\\
Algorithm 2:
\begin{enumerate}
\item Draw all connected one-loop diagrams with $k$ vertices, labelled by $u_1$, $u_2$, ..., $u_k$, including diagrams with self-loops.
\item Select one edge $e_C$, such that upon removal of this edge the diagram becomes a connected tree diagram.
\item For every edge $e \neq e_C$ connecting $u_k$ and $u_l$ use an operation $U_{ij}$, where $i$ labels a quark-antiquark pair in $u_k$ and
$j$ labels a quark-antiquark pair in $u_l$ and sum over $i$ and $j$.
The order in which these operations are performed is irrelevant, since the operations for a given graph are associative.
\item If $e_C$ is not a self-loop and connects $u_k$ and $u_L$ use an operation $C_{ij}$, where $i$ labels a quark-antiquark pair in $u_k$ and
$j$ labels a quark-antiquark pair in $u_l$ and sum over $i$ and $j$.
\item If $e_C$ is a self-loop connected to $u_k$ use an operation $C_{ij}$, where $i$ and $j$ label quark-antiquark pairs in $u_k$ with $i \le j$
and sum over $i$ and $j$ subject to $i \le j$.
\item Divide by the symmetry factor of the diagram.
\item Sum over all diagrams.
\end{enumerate}
A few comments on this algorithm:
\begin{description}
\item{-} Note that the minimal number of vertices is $1$. In this case there is only one one-loop diagram.
This diagram consists of one vertex with a self-loop (tadpole). Step 3 and 4 are not relevant in this case.

\item{-} If $e_C$ is a self-loop the case $i=j$ is included in step 5, which corresponds to the operation $C_{ii}$, or in other
words to a loop closing operation on the same quark line.

\item{-} A non-trivial symmetry factor of two is obtained if the vertices $u_k$ and $u_l$ are connected by two edges.
In this case one of the two edges must be selected as the special edge $e_C$.
\end{description}

\section{The method}
\label{sect:method}

We consider amplitudes with $n_g$ external gluons, $n_q$ external quarks and $n_q$ external antiquarks.
The total number of external particles is then
\bq
 n & = & n_g + 2 n_q.
\eq
We label the external quarks by
\bq
 q_1, q_2, ..., q_{n_q}
\eq
and the external antiquarks by
\bq
 \bar{q}_1, \bar{q}_2, ..., \bar{q}_{n_q}.
\eq
We assume that $q_i$ has the same flavour as $\bar{q}_i$ and that the flavours of all quark-antiquark pairs are distinct,
in other words
\bq
 \delta^{\mathrm{flav}}_{\bar{q}_i q_j } & = & \delta_{ij}.
\eq
In the following we discuss the decomposition of partial amplitudes into primitive amplitudes.
Partial amplitudes are the coefficients of colour factors.
We will show that the information contained in the colour factors determines the decomposition into primitive amplitudes.
We first translate the information contained in the colour factors into cyclic words.
For closed strings this is simple: Each closed string defines a cyclic word $v$:
\bq
 c_{\mathrm{closed}}(g_1,g_2,...,g_n) 
 & \Rightarrow &
 v = (g_1 g_2 ... g_n).
\eq
For open strings the situation is a little bit more complicated.
Consider the product of open strings appearing in eq.~(\ref{tree_multi_quark}) and eq.~(\ref{one_loop_multi_quark})
\bq
 & &
 c_{\mathrm{open}}\left(q_1,g_{\sigma_1},...,g_{\sigma_{i_1}},\bar{q}_{\pi_1}\right)
 c_{\mathrm{open}}\left(q_2,g_{\sigma_{i_1+1}},...,g_{\sigma_{i_1+i_2}},\bar{q}_{\pi_2}\right)
 \nonumber \\
 & & 
 ...
 c_{\mathrm{open}}\left(q_{n_q},g_{\sigma_{i_1+...+i_{n_q-1}+1}},...,g_{\sigma_{i_1+...+i_{n_q}}},\bar{q}_{\pi_{n_q}}\right)
\eq
with $n_g=i_1+...+i_{n_q}$.
The information is encoded in a permutation $\sigma \in S_{n_g}$ of the gluons, a permutation $\pi \in S_{n_q}$ of the 
antiquarks and a partition of $n_g$ into non-negative integers $\{i_1,...,i_{n_q}\}$ corresponding to the 
distribution of $n_g$ gluons among $n_q$ open strings.
We now focus on the permutation $\pi$ of the antiquarks.
Suppose that $\pi$ consists of $r$ cycles and that the order of the $i$-th cycle is $k_i$.
Clearly
\bq
 k_1 + k_2 + ... + k_r & = & n_q.
\eq
Without loss of generality we can relabel the quarks and antiquarks such that
\bq
 \pi & = & 
 \left(1,2,...,k_1\right) 
 \left(k_1+1,...,k_1+k_2\right)
 ...
 \left(k_1+...+k_{r-1}+1,...,k_1+...+k_r\right).
\eq
In order to keep the notation simple, we also assume that we have relabelled the gluons such that
\bq
 \sigma & = & (1,2,...,n_g).
\eq
The colour factor corresponding to the $j$-th cycle is
\bq
\label{cycle_colour_factor}
 & &
 c_{\mathrm{open}}\left( q_{k_1+...+k_{j-1}+1}, g_{i_1+i_2+...+i_{k_1+...+k_{j-1}}+1}, ..., g_{i_1+...+i_{k_1+...+k_{j-1}}+i_{k_1+...+k_{j-1}+1}}, \bar{q}_{k_1+...+k_{j-1}+2} \right)
 \nonumber \\
 & &
 c_{\mathrm{open}}\left( q_{k_1+...+k_{j-1}+2}, g_{i_1+i_2+...+i_{k_1+...+k_{j-1}+1}+1}, ..., g_{i_1+...+i_{k_1+...+k_{j-1}+2}}, \bar{q}_{k_1+...+k_{j-1}+3} \right)
 \nonumber \\
 & &
 ...
 c_{\mathrm{open}}\left( q_{k_1+...+k_{j-1}+k_j}, g_{i_1+i_2+...+i_{k_1+...+k_{j-1}+k_j-1}+1}, ..., g_{i_1+...+i_{k_1+...+k_{j-1}+k_j}}, \bar{q}_{k_1+...+k_{j-1}+1} \right).
 \nonumber \\
\eq
This colour factor contains $k_j$ open strings.
Eq.~(\ref{cycle_colour_factor}) defines a cyclic word
\bq
\label{def_u_j}
 u_j & = &
 \left( q_{k_1+...+k_{j-1}+1} ... \bar{q}_{k_1+...+k_{j-1}+2}
 q_{k_1+...+k_{j-1}+2} ... \bar{q}_{k_1+...+k_{j-1}+3}
 \right.
 \nonumber \\
 & &
 \left.
 ...
 q_{k_1+...+k_{j-1}+k_j} ... \bar{q}_{k_1+...+k_{j-1}+1} \right).
\eq
The cyclic word is obtained from the colour factor by concatenating the arguments of the open strings such that a quark follows immediately after the corresponding antiquark
in the cyclic order.
In this way we obtain from the $n_q$ open strings $r$ cyclic words $u_1$, $u_2$, ..., $u_r$.
We point out that in general an open string alone does not define a cyclic word, it is the product of open strings corresponding to a cycle of the permutation $\pi$ which defines a
cyclic word.

We are now in a position to present the decomposition of partial amplitudes into primitive amplitudes.
We distinguish the following six cases: 
The first two cases are related to tree amplitudes. 
We start with amplitudes with no external quarks (section~\ref{subsect:tree_gluon})
and then discuss amplitudes with external quarks (section~\ref{subsect:tree_quark}).
We then proceed to one-loop amplitudes and consider first one-loop amplitudes with a closed fermion loop (i.e. the $N_f$-part).
Again we distinguish between amplitudes with no external quarks (section~\ref{subsect:loop_gluon_Nf})
and amplitudes with external quarks (section~\ref{subsect:loop_quark_Nf}).
Finally we consider one-loop amplitudes without a closed fermion loop. 
Also here we distinguish between amplitudes with no external quarks (section~\ref{subsect:loop_gluon})
and amplitudes with external quarks (section~\ref{subsect:loop_quark}).
In all six cases we show how to express the partial amplitudes in terms of the primitive ones.
The results for the all-gluon amplitudes (i.e. no external quarks) and the amplitudes with $\bar{q} q + (n-2)$ gluons are well known.
The results for the multi-quark case are new.
In section~\ref{subsect:loop_one_quark} we show how the formula for the multi-quark case reduces in the single quark case to the known decomposition.

\subsection{Tree amplitudes with no external quarks}
\label{subsect:tree_gluon}

If no external quarks are present ($n_q=0$), all external particles are gluons and we have $n=n_g$.
This case was already discussed in eq.~(\ref{example_partial_to_primitive}).
The colour factor $c_{\mathrm{closed}}(g_{\sigma_1}, g_{\sigma_2}, ..., g_{\sigma_n} )$ in eq.~(\ref{colour_decomp_pure_gluon_II}) 
defines a cyclic word $v=(g_{\sigma_1} g_{\sigma_2} ... g_{\sigma_n} )$ and we have for the associated partial amplitude
\bq
 A_n^{(0)} & = & P_n^{(0)}\left(v\right).
\eq
In this case the decomposition of the partial amplitude into primitive amplitudes is trivial and consists only of a single term.
The primitive amplitude $P_n^{(0)}(v)=P_n^{(0)}\left(g_{\sigma_1} g_{\sigma_2} ... g_{\sigma_n} \right)$ is calculated from planar cyclic ordered diagrams 
with the Feynman rules of appendix \ref{appendix:colour_ordered_rules}.

\subsection{Tree amplitudes with external quarks}
\label{subsect:tree_quark}

We now consider tree amplitudes with external quarks.
The colour decomposition of the full amplitude into partial amplitudes is given in eq.~(\ref{tree_multi_quark}).
The colour factors are specified by a permutation $\sigma \in S_{n_g}$ of the gluons, a permutation $\pi \in S_{n_q}$ of the 
antiquarks and a partition of $n_g$ into non-negative integers $\{i_1,...,i_{n_q}\}$ corresponding to the 
distribution of $n_g$ gluons among $n_q$ open strings.
Let us assume that the permutation $\pi$ consists of $r$ cycles.
We then have $r$ cyclic words 
\bq
 u_1, u_2, ..., u_r,
\eq
according to eq.~(\ref{def_u_j}).
Each cyclic word $u_j$ corresponds to a colour cluster. The individual colour clusters are connected by $U(1)$-gluons. 
Therefore the number of cycles of the permutation $\pi$ determines the number of $U(1)$-gluons in the amplitude:
If $\pi$ consists of $r$ cycles, the amplitude will contain $(r-1)$ gluons of type $U(1)$.
The decomposition of the partial amplitude into primitive amplitudes is given by
\bq
 A^{(0)}_n
 & = &
 \left(-\frac{1}{N}\right)^{r-1}
 \sum\limits_{w \in U\left(u_1, ..., u_r\right)} 
 P^{(0)}_n\left( w \right),
\eq
where the sum is over all cyclic words appearing in the shuffle operation $U(u_1, ..., u_r)$ defined in algorithm 1 in section~\ref{subsection:shuffle}.
The factor $(-1/N)^{r-1}$ in front comes from the Fierz identity in eq.~(\ref{fierz_identity}).

\subsection{One-loop amplitudes with a closed fermion loop and no external quarks}
\label{subsect:loop_gluon_Nf}

We now turn to one-loop amplitudes and consider first one-loop amplitudes with a closed fermion loop and no external quarks.
The decomposition of the all-gluon one-loop amplitude into partial amplitudes is given in eq.~(\ref{one_loop_all_gluon}).
The two closed strings $c_{\mathrm{closed}}(g_{\sigma_1},...,g_{\sigma_{n-m}})$ and $c_{\mathrm{closed}}(g_{\sigma_{n-m+1}}$, ..., $g_{\sigma_n})$
define two cyclic words $v_1=(g_{\sigma_1} ... g_{\sigma_{n-m}})$ and $v_2=(g_{\sigma_{n-m+1}} ... g_{\sigma_n})$.
Amplitudes with $n=n_g$ external gluons and a closed fermion loop have just a single colour factor and correspond therefore to $m=0$.
Thus
\bq
 v_1=\left(g_{\sigma_1} ... g_{\sigma_{n}}\right)
 & \mbox{and} &
 v_2=().
\eq
All $m=0$ amplitudes are primitive and we have
\bq
  A^{(1) [1/2]}_{n,0}
 & = &
  \frac{N_f}{N} P^{(1) [1/2]}_{n}\left( v_1 \right).
\eq
The factor $N_f$ corresponds to the $N_f$ possible flavours in the loop, the factor $1/N$ compensates
the second trace $c_{\mathrm{closed}}()=N$ in eq.~(\ref{one_loop_all_gluon}), which is not present in the $N_f$-part.
$P^{(1) [1/2]}_{n}$ is computed from planar cyclic ordered one-loop diagrams with a closed fermion loop.
The arrow of the fermion line is counter-clockwise.
In the computation of $P^{(1) [1/2]}_{n}$ a minus sign for a closed fermion loop has to be taken into account.

\subsection{One-loop amplitudes with a closed fermion loop and external quarks}
\label{subsect:loop_quark_Nf}

We now consider one-loop amplitudes with a closed fermion loop and with external quarks.
The colour decomposition of the full amplitude into partial amplitudes is given in eq.~(\ref{one_loop_multi_quark}).
The colour factors of the partial amplitudes are specified by a permutation $\sigma \in S_{n_g}$ of the gluons, a permutation $\pi \in S_{n_q}$ of the 
antiquarks and a partition of $n_g$ into non-negative integers $\{i_1,...,i_{n_q},m\}$ corresponding to the 
distribution of $n_g$ gluons among $n_q$ open strings and one closed string. 
Let us assume that the permutation $\pi$ consists of $r$ cycles.
We then have $r$ cyclic words 
\bq
 u_1, u_2, ..., u_r,
\eq
according to eq.~(\ref{def_u_j}).
The closed string defines an additional cyclic word
\bq
 v & = & \left( g_{\sigma_{i_1+...+i_{n_q}+1}} ... g_{\sigma_{n_g}} \right).
\eq
The closed string may be empty.
If the permutation $\pi$ consists of $r$ cycles, we may have either $r$ or $(r-1)$ gluons of type $U(1)$ in the amplitude.
If we have $r$ gluons of type $U(1)$, then the closed fermion loop is colour disconnected from all external quark lines.
Only in this case we may have $m>0$.
It should be mentioned that $m>0$ is not required, $m=0$ is also allowed in this case.
On the other hand in the case of $(r-1)$ gluons of type $U(1)$ the closed fermion loop is necessarily colour-connected to one of the colour clusters defined
by the cyclic words $u_1$, $u_2$, ..., $u_r$.
In this case we must have $m=0$.
We use a superscript $[1/2]$ as in
\bq
 u_j^{[1/2]}
\eq
to indicate the word colour-connected to the closed fermion loop.
All external quark lines in $u_j^{[1/2]}$ are assigned a left-moving routing label.

The decomposition of the partial amplitude into primitive amplitudes reads
\bq
\label{nf_decomposition}
 A^{(1) [1/2]}_{n,m}
 & = &
 \delta_{m,0}
 \frac{N_f}{N}
 \left(-\frac{1}{N}\right)^{r-1}
 \sum\limits_{j=1}^{r}
 \;\;
 \sum\limits_{w \in U\left( u_1, ..., u_{j-1}, u_j^{[1/2]}, u_{j+1}, ...,  u_{r}\right)} 
 P^{(1) [1/2]}_n\left( w \right)
 \nonumber \\
 & &
 +
 N_f
 \left(-\frac{1}{N}\right)^{r}
 \sum\limits_{w \in U\left(u_1, ..., u_{r}, v \right)} 
 P^{(1) [1/2]}_n\left( w \right).
\eq
The first line corresponds to the case, where the closed fermion loop is colour-connected to one of the colour clusters.
The Kronecker $\delta_{m,0}$ ensures that this term contributes only for $m=0$.
The factor $N_f$ corresponds to the $N_f$ possible flavours in the loop while the factor $1/N$ compensates
the trace $c_{\mathrm{closed}}()=N$ in eq.~(\ref{one_loop_multi_quark}), which is not present in this case.
The factor $(-1/N)^{r-1}$ comes from the Fierz identity in eq.~(\ref{fierz_identity}) and corresponds to $(r-1)$ gluons of type $U(1)$.

The second line corresponds to the case, where the closed fermion loop is not colour-connected to any external quark line.
Note that in this case the factor $N_f$ is not accompanied by a factor $1/N$.
(There is no fake empty closed string to be divided out.)
The factor $(-1/N)^r$ comes from the Fierz identity in eq.~(\ref{fierz_identity}) and corresponds to $r$ gluons of type $U(1)$.

We remark that in all cases in the computation of $P^{(1) [1/2]}_{n}$ a minus sign for a closed fermion loop has to be taken into account.

\subsection{One-loop amplitudes with no closed fermion loop and no external quarks}
\label{subsect:loop_gluon}

In this case all external particles are gluons ($n=n_g$) and the particles in the loop are either gluons or ghosts.
The colour decomposition of the all-gluon one-loop amplitude into partial amplitudes is given in eq.~(\ref{one_loop_all_gluon}).
The two closed strings $c_{\mathrm{closed}}(g_{\sigma_1},...,g_{\sigma_{n-m}})$ and $c_{\mathrm{closed}}(g_{\sigma_{n-m+1}},...,g_{\sigma_n})$
define two cyclic words 
\bq
 v_1=\left(g_{\sigma_1} ... g_{\sigma_{n-m}}\right),
 & \mbox{and} &
 v_2=\left(g_{\sigma_{n-m+1}} ... g_{\sigma_n}\right).
\eq
The relation between partial amplitudes and primitive amplitudes is well known \cite{Bern:1990ux} and we have
\bq
 A^{(1) [1]}_{n,m}
 & = & (-1)^m
 \sum\limits_{w \in v_1 \circledcirc v_2^T}  
 P^{(1) [1]}_{n}\left( w \right).
\eq

\subsection{One-loop amplitudes with no closed fermion loop and external quarks}
\label{subsect:loop_quark}

The case of one-loop amplitudes with no closed fermion loops and external quarks is the most complicated one.
On the one hand we can have the double-ring structure already encountered in the case of a pure gluon one-loop amplitude.
In the case of external quarks, the external quarks can be attached a priori to either colour ring.
On the other hand there can be $U(1)$-gluons, as in the case of tree amplitudes with external quarks.

The colour decomposition of the full amplitude into partial amplitudes is given in eq.~(\ref{one_loop_multi_quark}).
The colour factors of the partial amplitudes are specified by a permutation $\sigma \in S_{n_g}$ of the gluons, a permutation $\pi \in S_{n_q}$ of the 
antiquarks and a partition of $n_g$ into non-negative integers $\{i_1,...,i_{n_q},m\}$ corresponding to the 
distribution of $n_g$ gluons among $n_q$ open strings and one closed string. 
Let us assume that the permutation $\pi$ consists of $r$ cycles.
We then have $r$ cyclic words 
\bq
 u_1, u_2, ..., u_r,
\eq
according to eq.~(\ref{def_u_j}).
The closed string defines an additional cyclic word
\bq
 v & = & \left( g_{\sigma_{i_1+...+i_{n_q}+1}} ... g_{\sigma_{n_g}} \right).
\eq
The closed string may be empty.
We denote by $p$ the number of $U(1)$-gluons.
We will first show that the number of $U(1)$-gluons $p$ is either $r$, $(r-1)$ or $(r-2)$.
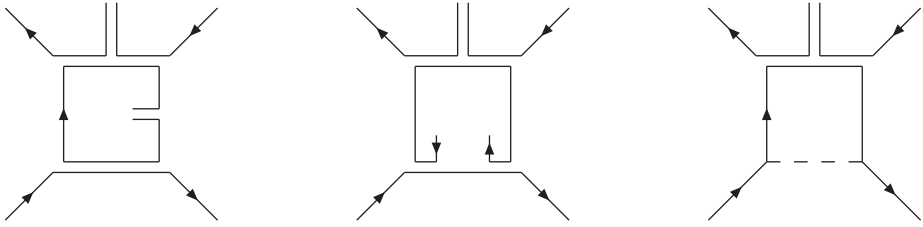
\begin{figure}
\begin{center}
\begin{picture}(130,100)(0,0)
\ArrowLine(28,72)(10,90)
\ArrowLine(10,10)(28,28)
\ArrowLine(72,28)(90,10)
\ArrowLine(90,90)(72,72)
\ArrowLine(32,32)(32,68)
\Line(32,68)(68,68)
\Line(68,68)(68,52)
\Line(68,48)(68,32)
\Line(68,32)(32,32)
\Line(72,72)(52,72)
\Line(48,72)(28,72)
\Line(28,28)(72,28)
\Line(48,72)(48,92)
\Line(52,72)(52,92)
\Line(68,48)(58,48)
\Line(68,52)(58,52)
\end{picture} 
\begin{picture}(130,100)(0,0)
\ArrowLine(28,72)(10,90)
\ArrowLine(10,10)(28,28)
\ArrowLine(72,28)(90,10)
\ArrowLine(90,90)(72,72)
\Line(32,32)(32,68)
\Line(32,68)(68,68)
\Line(68,68)(68,32)
\Line(68,32)(60,32)
\Line(40,32)(32,32)
\ArrowLine(40,42)(40,32)
\ArrowLine(60,32)(60,42)
\Line(72,72)(52,72)
\Line(48,72)(28,72)
\Line(28,28)(72,28)
\Line(48,72)(48,92)
\Line(52,72)(52,92)
\end{picture} 
\begin{picture}(130,100)(0,0)
\ArrowLine(28,72)(10,90)
\ArrowLine(10,10)(32,32)
\ArrowLine(68,32)(90,10)
\ArrowLine(90,90)(72,72)
\ArrowLine(32,32)(32,68)
\Line(32,68)(68,68)
\Line(68,68)(68,32)
\DashLine(68,32)(32,32){5}
\Line(72,72)(52,72)
\Line(48,72)(28,72)
\Line(48,72)(48,92)
\Line(52,72)(52,92)
\end{picture} 
\caption{\label{figure_one_loop_1}
Examples of colour-flow diagrams for the three cases discussed in the text.
The first diagram shows an example with a closed string corresponding to case 1.
The second diagram shows an example without a closed string.
Here, the particles attached to the outer ring are colour-disconnected from particles attached to the inner ring (case 2).
The third diagram shows also an example without a closed string.
Here, the particles attached to the outer ring are colour-connected to the particles attached to the inner ring (case 3).
}
\end{center}
\end{figure}
We can distinguish the following three cases:
\begin{description}
\item{Case 1:} 
There is a closed string with zero or more gluons attached. 
We have $m \ge 0$ and $p=r-1$.
This is the tree-level relation between the cycles $r$ of the permutation $\pi$ and the number of $U(1)$-gluons.
The inner ring of the loop contains only gluons and does not affect this relation.
In this case we have the cyclic words $u_1$, $u_2$, ..., $u_r$ and the cyclic word $v$.
One of the cyclic words $u_1$, $u_2$, ..., $u_r$ corresponds to a loop structure, while the others correspond to tree structures.
In the case where the cyclic word $u_j$ corresponds to a loop structure we indicate this with a superscript $[1]$ as in
\bq
 u_j^{[1]}.
\eq
All external quark lines in $u_j^{[1]}$ are assigned a left-moving routing label.
\item{Case 2:} 
There is no closed string, but particles attached to the outer ring are colour-disconnected from particles attached to the inner ring.
(This requires at least two quark-antiquark pairs, one pair attached to the outer ring and another pair attached to the inner ring.)
We have $m=0$ and $p=r-2$.
In this case we have the cyclic words $u_1$, $u_2$, ..., $u_r$, while the cyclic word $v$ is empty.
The loop is made out of two cyclic words, say $u_i$ and $u_j$ (one for the outer ring, one for the inner ring), while all other cyclic words correspond
to tree-like structures.
\item{Case 3:} 
There is no closed string and particles attached to the outer ring are colour-connected to particles attached to the inner ring.
We have $m=0$ and $p=r$. In this case there is at least one $U(1)$-gluon propagating in the loop.
We have the cyclic words $u_1$, $u_2$, ..., $u_r$, while the cyclic word $v$ is empty.
\end{description}
Only in the first case we can have $m \ge 0$, in all other cases we have $m=0$.
Examples of colour-flow diagrams corresponding to the three cases are shown in fig.~(\ref{figure_one_loop_1}).
The decomposition of the partial amplitude into primitive amplitudes reads
\bq
\label{decomposition}
 A^{(1) [1]}_{n,m}
 & = &
 \left(-1\right)^m
 \left(-\frac{1}{N}\right)^{r-1}
 \sum\limits_{j=1}^r
 \;\;
 \sum\limits_{w \in U\left( u_1, ..., u_{j-1}, u_j^{[1]}, u_{j+1}, ..., u_{r} \right) \circledcirc v^T} 
 P^{(1) [1]}_{n}\left(w\right)
 \nonumber \\
 & &
 +
 \delta_{m,0}
 \frac{1}{N}
 \left(-\frac{1}{N}\right)^{r-2}
 \;\;
 \sum\limits_{i=1}^{r-1}
 \sum\limits_{j=i+1}^r
 \left(-1\right)^{n_{u_j}}
 \sum\limits_{w \in U\left( u_1, ..., \widehat{u_i}, ..., \widehat{u_j}, ..., u_{r}, \left( u_i \circledcirc u_j^T \right)^{[1]} \right)} 
 P^{(1) [1]}_{n}\left(w\right)
 \nonumber \\
 & &
 +
 \delta_{m,0}
 \frac{1}{N}
 \left(-\frac{1}{N}\right)^{r}
 \;\;
 \sum\limits_{w \in CU\left(u_1, ..., u_{r}\right)} 
 P^{(1) [1]}_{n}\left(w\right).
\eq
The first line corresponds to the case 1. 
The factor $(-1/N)^{r-1}$ comes from the Fierz identity for $(r-1)$ gluons of type $U(1)$.
The second line corresponds to the case 2.
The Kronecker $\delta_{m,0}$ ensures that this term contributes only for $m=0$.
The factor $1/N$ compensates the trace $c_{\mathrm{closed}}()=N$, which is not present in this case.
The factor $(-1/N)^{r-2}$ comes from the Fierz identity for $(r-2)$ gluons of type $U(1)$.
The notation $\widehat{u_i}$ indicates that this term is omitted.
$n_{u_j}$ denotes the length of the cyclic word $u_j$.
The quark lines in the term $( u_i \circledcirc u_j^T )^{[1]}$ have routing labels assigned as follows:
All quark lines of $u_i$ are left movers, while all quark lines of $u_j^T$ are right movers.
The third line corresponds to the case 3.
Again, the Kronecker $\delta_{m,0}$ ensures that this term contributes only for $m=0$ and
the factor $1/N$ compensates the trace $c_{\mathrm{closed}}()=N$, which is not present in this case.
The factor $(-1/N)^{r}$ comes from the Fierz identity for $r$ gluons of type $U(1)$.

\subsection{One-loop amplitudes with one external quark pair}
\label{subsect:loop_one_quark}

In this section we discuss specifically one-loop amplitudes with one external quark pair $\bar{q} q$ and $(n-2)$ gluons.
We show how the formulae eq.~(\ref{decomposition}) and (\ref{nf_decomposition}) reduce for the case of one external quark pair
to the known decomposition of the partial amplitudes into primitive amplitudes.

We label the external particles by $q$, $\bar{q}$ and $g_1$, $g_2$, ..., $g_{n-2}$.
The quark permutation $\pi$ is trivial and we have $r=1$.
As cyclic words we can have
\bq
 u_1=\left(q g_1 ... g_{n-m-2} \bar{q}\right),
 & &
 v=\left(g_{n-m-1} ... g_{n-2}\right),
 \;\;\;\;\;\;
 0 \le m \le n-2.
\eq
In the case $m=0$ the cyclic word $v$ is empty.
We discuss this case first.
For $m=0$ we have $u_1=(q g_1 ... g_{n-2} \bar{q})$ and $v=()$.
The eqs.~(\ref{decomposition}) and (\ref{nf_decomposition}) yield
\bq
\label{compare_case_m_zero}
 A_{n,0}^{(1)[1]} & = &
 P_n^{(1)[1]}\left(q^L g_1 ... g_{n-2} \bar{q}^L\right)
 - \frac{1}{N^2} P_n^{(1)[1]}\left(q^R g_1 ... g_{n-2} \bar{q}^R \right),
 \\
 A_{n,0}^{(1)[1/2]} & = &
 \frac{N_f}{N} P_n^{(1)[1/2]}\left(q^L g_1 ... g_{n-2} \bar{q}^L\right)
 - \frac{N_f}{N} \left(-1\right)^{n-2} P_n^{(1)[1/2]}\left(q^L \bar{q}^L g_{n-2} ... g_2 g_1 \right).
 \nonumber 
\eq
The second term in the $N_f$-part corresponds to tadpoles and vanishes for massless particles in dimensional regularisation.
Taken this into account, eq.~(\ref{compare_case_m_zero}) agrees with eq.~(4.1) of \cite{Bern:1994fz}.

Let us now discuss the case $m>0$. We have two non-trivial cyclic words
\bq
u_1= \left(q g_1 ... g_{n-m-2} \bar{q} \right) 
& \mbox{and} &
v= \left(g_{n-m-1} ... g_{n-2} \right).
\eq
The eqs.~(\ref{decomposition}) and (\ref{nf_decomposition}) yield
\bq
\label{compare_case_m_not_zero}
 A_{n,m}^{(1)[1]} & = &
 \left(-1\right)^m
 \sum\limits_{w \in (q^L g_1 ... g_{n-m-2} \bar{q}^L) \circledcirc (g_{n-2} ... g_{n-m-1})}
 P_n^{(1)[1]}\left(w\right),
 \\
 A_{n,m}^{(1)[1/2]} & = &
 - \frac{N_f}{N} 
 \left(-1\right)^{n-m-2}
 \sum\limits_{w \in (q^L \bar{q}^L g_{n-m-2} ... g_1) \circledcirc (g_{n-m-1} ... g_{n-2})}
 P_n^{(1)[1/2]}\left(w\right).
 \nonumber 
\eq
The $[1]$-part agrees directly with eq.~(4.4) of \cite{Bern:1994fz}.
Using the reflection identity eq.~(\ref{reflection_identity}) we can show that the $[1/2]$-part is equivalent to
\bq
\label{using_reflection_identity}
 A_{n,m}^{(1)[1/2]} & = &
 - \frac{N_f}{N} 
 \left(-1\right)^{m}
 \sum\limits_{w \in (q^R g_1 ... g_{n-m-2} \bar{q}^R) \circledcirc (g_{n-2} ... g_{n-m-1})}
 P_n^{(1)[1/2]}\left(w\right),
\eq
which in turn agrees with eq.~(4.4) of \cite{Bern:1994fz}.
We remind the reader that the decomposition of partial amplitudes into primitive amplitudes is in general not unique.
The two decompositions of the $N_f$-part in eq.~(\ref{compare_case_m_not_zero}) and eq.~(\ref{using_reflection_identity}) are a manifestation of this.

\section{Examples}
\label{sect:examples}

In this section we give a few examples to illustrate our method.
In the first example we discuss the one-loop amplitude ${\cal A}^{(1)}_4(q_1,\bar{q}_1,q_2,\bar{q}_2)$.
This is the next-to-complicated case beyond the amplitudes with one quark-antiquark pair discussed in section~(\ref{subsect:loop_one_quark})
and illustrates already most of the features of our method.

As a second example we consider the one-loop amplitude ${\cal A}^{(1)}_6(q_1,\bar{q}_1,q_2,\bar{q}_2,g_1,g_2)$, involving two gluons in addition to two
quark-antiquark pairs.
In this example we give the decomposition of two selected partial amplitudes into primitive amplitudes.

All our examples can be worked out on the back of an envelope. This should be compared to other methods, which usually resort to computers for the 
six-parton case.
Of course, our method is also suited for an automated computer implementation and the examples below serve only as an illustration of our algorithm.

\subsection{An example with four quarks}

We start with the one-loop amplitude ${\cal A}^{(1)}_4(q_1,\bar{q}_1,q_2,\bar{q}_2)$.
In this case there are two possible colour factors, given by
\bq
\label{colour_factors_4q}
 N \; \delta_{i_{q_1} j_{\bar{q}_2}} \delta_{i_{q_2} j_{\bar{q}_1}}
 & \mbox{and} &
 N \; \delta_{i_{q_1} j_{\bar{q}_1}} \delta_{i_{q_2} j_{\bar{q}_2}}.
\eq
The first one corresponds to $\pi=(1,2)$, the second one to $\pi=(1)(2)$.
The factor $N$ corresponds to an empty closed string.
We discuss the two cases separately.

\subsubsection{The case $\pi=(1,2)$}

Here we have $r=1$, $m=0$ and the cyclic word
\bq
 u_1 & = & \left(q_1 \bar{q}_2 q_2 \bar{q}_1 \right),
\eq
$v=()$ is empty and contributes a factor $N$ to the colour factor in eq.~(\ref{colour_factors_4q}).
We start with the non-$N_f$-part.
In eq.~(\ref{decomposition}) only the first and the third line contribute, the second line would require two cyclic words $u_i$ and $u_j$.
Thus
\bq
 A_{4,0}^{(1)[1]}
 & = &
 P_4^{(1)[1]}\left(u_1^{[1]}\right)
 - \frac{1}{N^2} \sum\limits_{w\in CU\left(u_1\right)}P_4^{(1)[1]}\left(w\right).
\eq
We work out the second term.
We have
\bq
 CU\left(u_1\right) & = &
 C_{11}\left(u_1\right) + C_{12}\left(u_1\right) + C_{22}\left(u_1\right),
\eq
and 
\bq
 C_{11}\left(u_1\right) 
 & = &
 \left( q_1^R \bar{q}_2^L q_2^L \bar{q}_1^R \right),
 \nonumber \\
 C_{22}\left(u_1\right)
 & = &
 \left( q_1^L \bar{q}_2^R q_2^R \bar{q}_1^L \right),
 \nonumber \\
 C_{12}\left(u_1\right)
 & = &
 \left( q_1^L \bar{q}_2^L q_2^L \bar{q}_1^L \right)
 +
 \left( q_1^L q_2^R \bar{q}_2^R \bar{q}_1^L \right)
 +
 \left( q_1^R \bar{q}_1^R \bar{q}_2^L q_2^L \right)
 +
 \left( q_1^R \bar{q}_1^R q_2^R \bar{q}_2^R \right).
\eq
We can use the reflection identity in eq.~(\ref{reflection_identity})
to convert any primitive amplitude with $q_1^R$ to a primitive amplitude with $q_1^L$, for example
\bq
 P_4^{(1)[1]}\left( q_1^R \bar{q}_1^R \bar{q}_2^L q_2^L \right)
 & = &
 P_4^{(1)[1]}\left( q_1^L q_2^R \bar{q}_2^R \bar{q}_1^L \right).
\eq
We arrive at
\bq
\label{eq1}
 A_{4,0}^{(1)[1]}
 & = &
 \left( 1- \frac{2}{N^2} \right)
 P_4^{(1)[1]}\left(q_1^L \bar{q}_2^L q_2^L \bar{q}_1^L \right)
 \\
 & &
 - \frac{2}{N^2}P_4^{(1)[1]}\left(q_1^L q_2^R \bar{q}_2^R \bar{q}_1^L \right)
 - \frac{1}{N^2}P_4^{(1)[1]}\left(q_1^L \bar{q}_2^R q_2^R \bar{q}_1^L \right)
 - \frac{1}{N^2}P_4^{(1)[1]}\left(q_1^L \bar{q}_1^L q_2^R \bar{q}_2^R \right).
 \nonumber
\eq
We note that
\bq
\label{side_relation}
 P_4^{(1)[1]}\left(q_1^L \bar{q}_1^L q_2^R \bar{q}_2^R \right)
 & = & 
 -
 P_4^{(1)[1]}\left(q_1^L \bar{q}_1^L \bar{q}_2^L q_2^L \right)
\eq
and with this relation our result is equivalent to \cite{Ita:2011ar}.

We now discuss the $N_f$-part. Eq.~(\ref{nf_decomposition}) gives
\bq
 A_{4,0}^{(1)[1/2]}
 & = &
 \frac{N_f}{N}
 P_4^{(1)[1/2]}\left(u_1^{[1/2]}\right)
 - \frac{N_f}{N} \sum\limits_{w\in U\left(u_1,v\right)}P_4^{(1)[1/2]}\left(w\right).
\eq
The second term corresponds to tadpoles. These vanish in massless QCD in dimensional regularisation.
We will ignore them in the following.
Thus
\bq
 A_{4,0}^{(1)[1/2]}
 & = &
 \frac{N_f}{N}
 P_4^{(1)[1/2]}\left(q_1^L \bar{q}_2^L q_2^L \bar{q}_1^L \right).
\eq
This agrees with the result of \cite{Ita:2011ar}.

\subsubsection{The case $\pi=(1)(2)$}

We now discuss the second colour factor corresponding to $\pi=(1)(2)$.
We have $r=2$, $m=0$ and
\bq
 u_1=\left(q_1\bar{q}_1\right),
 & &
 u_2=\left(q_2\bar{q}_2\right),
\eq
$v=()$ is empty and contributes a factor $N$ to the colour factor in eq.~(\ref{colour_factors_4q}).
In the non-$N_f$-part we obtain now from eq.~(\ref{decomposition})
\bq
 A_{4,0}^{(1)[1]}
 & = &
 - \frac{1}{N} \sum\limits_{w \in U\left(u_1^{[1]},u_2\right)+U\left(u_1,u_2^{[1]}\right)} P_4^{(1)[1]}\left( w \right)
 \nonumber \\
 & &
 + \frac{1}{N} \sum\limits_{w \in \left(u_1\circledcirc u_2^T \right)^{[1]}} P_4^{(1)[1]}\left( w \right)
 + \frac{1}{N^3} \sum\limits_{w \in CU\left(u_1,u_2\right)} P_4^{(1)[1]}\left( w \right).
\eq
Now we work out the individual pieces. For the first term we need
\bq
 U\left(u_1^{[1]},u_2\right) = \left( q_1^L \bar{q}_1^L q_2^R \bar{q}_2^R \right),
 & &
 U\left(u_1,u_2^{[1]}\right) = \left( q_1^R \bar{q}_1^R q_2^L \bar{q}_2^L \right).
\eq
For the second term we work out
\bq
\label{crossed_lines_removed}
 \left(u_1\circledcirc u_2^T \right)^{[1]}
 & = &
 \left( q_1^L \bar{q}_2^R q_2^R \bar{q}_1^L \right)
 +
 \left( q_1^L \bar{q}_1^L q_2^R \bar{q}_2^R \right)
 +
 \left( q_1^L q_2^R \bar{q}_2^R \bar{q}_1^L \right).
\eq
Note that the cyclic word $( q_1^L \bar{q}_1^L \bar{q}_2^R q_2^R )$ has crossed fermion lines and does not appear in eq.~(\ref{crossed_lines_removed}).
We further have
\bq
\label{eq_with_symmetry_factor}
 CU\left(u_1,u_2\right)
 & = &
 CU_{11}\left(q_1\bar{q}_1q_2\bar{q}_2\right)
 +
 CU_{22}\left(q_1\bar{q}_1q_2\bar{q}_2\right)
 +
 \frac{1}{2} CU_{12}\left(q_1\bar{q}_1q_2\bar{q}_2\right).
\eq
$CU_{11}$ corresponds to a diagram, where $u_1$ and $u_2$ are connected by an edge and a self-loop attached to $u_1$.
In a similar way, $CU_{22}$ corresponds to a diagram, where $u_1$ and $u_2$ are connected by an edge with a self-loop attached to $u_2$.
On the other hand, $CU_{12}$ corresponds to a diagram, where $u_1$ and $u_2$ are connected by two edges.
This diagram has a symmetry factor of two, explaining the $1/2$ in front of $CU_{12}$ in eq.~(\ref{eq_with_symmetry_factor}).
We have
\bq
 CU_{11}\left(q_1\bar{q}_1q_2\bar{q}_2\right)
 & = &
 \left( q_1^R \bar{q}_2^L q_2^L \bar{q}_1^R \right),
 \nonumber \\
 CU_{22}\left(q_1\bar{q}_1q_2\bar{q}_2\right)
 & = &
 \left( q_1^L \bar{q}_2^R q_2^R \bar{q}_1^L \right),
 \nonumber \\
 CU_{12}\left(q_1\bar{q}_1q_2\bar{q}_2\right)
 & = & 
 \left( q_1^R \bar{q}_1^R q_2^R \bar{q}_2^R \right)
 +
 \left( q_1^R \bar{q}_1^R \bar{q}_2^L q_2^L \right)
 +
 \left( \bar{q}_1^L q_1^L q_2^R \bar{q}_2^R \right)
 +
 \left( \bar{q}_1^L q_1^L \bar{q}_2^L q_2^L \right).
\eq
We can again use the reflection identity in eq.~(\ref{reflection_identity}) to ensure that the quark line $1$ is left-moving.
We arrive at
\bq
 A_{4,0}^{(1)[1]}
 & = &
 \left( \frac{1}{N} + \frac{1}{N^3} \right) P_4^{(1)[1]}\left( q_1^L q_2^R \bar{q}_2^R \bar{q}_1^L \right)
 \nonumber \\
 & &
 + \frac{1}{N^3}
 \left[
 P_4^{(1)[1]}\left( q_1^L \bar{q}_1^L q_2^R \bar{q}_2^R \right)
 +
 P_4^{(1)[1]}\left( q_1^L \bar{q}_2^R q_2^R \bar{q}_1^L \right)
 +
 P_4^{(1)[1]}\left( q_1^L \bar{q}_2^L q_2^L \bar{q}_1^L \right)
 \right].
\eq
With the help of eq.~(\ref{side_relation}) one establishes the equivalence with \cite{Ita:2011ar}.

Let us now consider the $N_f$-part. Eq.~(\ref{nf_decomposition}) gives
\bq
 A_{4,0}^{(1)[1/2]}
 & = &
 - \frac{N_f}{N^2} \sum\limits_{w \in U\left(u_1^{[1/2]},u_2\right)+U\left(u_1,u_2^{[1/2]}\right)} P_4^{(1)[1/2]}\left( w \right)
 + \frac{N_f}{N^2} \sum\limits_{w \in U\left(u_1,u_2,v\right)} P_4^{(1)[1/2]}\left( w \right).
 \;\;\;\;\;\;
\eq
We have
\bq
 U\left(u_1^{[1/2]},u_2\right) & = &
 U_{12}\left(u_1^{[1/2]},u_2\right)
 +
 U_{2,\mathrm{loop}}\left(u_2,u_1^{[1/2]}\right),
 \nonumber \\
 U\left(u_1,u_2^{[1/2]}\right)
 & = &
 U_{12}\left(u_1,u_2^{[1/2]}\right)
 + 
 U_{1,\mathrm{loop}}\left(u_1,u_2^{[1/2]}\right).
\eq
As before we will ignore tadpoles.
We notice that $U_{12}$ will produce only tadpoles, while
\bq
\label{only_one_survives}
 U_{1,\mathrm{loop}}\left(u_1,u_2^{[1/2]}\right)
 & = &
 U_{2,\mathrm{loop}}\left(u_2,u_1^{[1/2]}\right)
 \nonumber \\
 & = &
 \left( q_1^L \bar{q}_2^L q_2^L \bar{q}_1^L \right)
 +
 \left( q_1^L q_2^L \bar{q}_2^L \bar{q}_1^L \right)
 +
 \left( q_1^L \bar{q}_1^L \bar{q}_2^L q_2^L \right)
 +
 \left( q_1^L \bar{q}_1^L q_2^L \bar{q}_2^L \right).
\eq
The last term has crossed fermion lines, the second and third term correspond to tadpoles, hence only the first term survives.

For $U\left(u_1,u_2,v\right)$ (with $v=()$ the empty word) we have three contributions
\bq
 U_{1,\mathrm{loop}}(U_{12}(u_1,u_2),v),
 \;\;\;\;\;\;
 U_{2,\mathrm{loop}}(U_{12}(u_1,u_2),v),
 \;\;\;\;\;\;
 U_{2,\mathrm{loop}}(u_2,U_{1,\mathrm{loop}}(u_1,v)).
\eq
Only the last one does not correspond to tadpoles.
We have
\bq
 U_{1,\mathrm{loop}}(u_1,v) & = & \left(\bar{q}_1^L q_1^L \right),
 \nonumber \\
 U_{2,\mathrm{loop}}\left(u_2,\left(\bar{q}_1^L q_1^L \right)\right))
 & = &
 \left( q_1^L \bar{q}_2^L q_2^L \bar{q}_1^L \right)
 +
 \left( q_1^L q_2^L \bar{q}_2^L \bar{q}_1^L \right)
 +
 \left( q_1^L \bar{q}_1^L \bar{q}_2^L q_2^L \right)
 +
 \left( q_1^L \bar{q}_1^L q_2^L \bar{q}_2^L \right).
\eq
Similar to the remark after eq.~(\ref{only_one_survives}) only the first term in the second equation survives.
Adding everything up, we obtain
\bq
 A_{4,0}^{(1)[1/2]}
 & = &
 -\frac{N_f}{N^2} P_4^{(1)[1/2]}\left( q_1^L \bar{q}_2^L q_2^L \bar{q}_1^L \right),
\eq
which agrees with \cite{Ita:2011ar}.

\subsection{An example with four quarks and two gluons}

As a second example we consider the one-loop amplitude ${\cal A}^{(1)}_6(q_1,\bar{q}_1,q_2,\bar{q}_2,g_1,g_2)$.
This example illustrates the effect of additional gluons.
\begin{table}
\begin{center}
\begin{tabular}{|c|l|c|c|l|}
\hline
 case & colour factor & $r$ & $m$ & cyclic words \\
 \hline
 $1$ & $N \; \delta_{i_{q_1} j_{g_1}} \delta_{i_{g_1} j_{g_2}} \delta_{i_{g_2} j_{\bar{q}_2}} \delta_{i_{q_2} j_{\bar{q}_1}}$ & $1$ & $0$ & $u_1=(q_1 g_1 g_2 \bar{q}_2 q_2 \bar{q}_1)$, $v=()$ \\
\hline
 $2$ & $N \; \delta_{i_{q_1} j_{g_1}} \delta_{i_{g_1} j_{\bar{q}_2}} \delta_{i_{q_2} j_{g_2}} \delta_{i_{g_2} j_{\bar{q}_1}}$ & $1$ & $0$ & $u_1=(q_1 g_1 \bar{q}_2 q_2 g_2 \bar{q}_1)$, $v=()$ \\
\hline
 $3$ & $\delta_{i_{q_1} j_{g_1}} \delta_{i_{g_1} j_{\bar{q}_2}} \delta_{i_{q_2} j_{\bar{q}_1}} \delta_{i_{g_2} j_{g_2}}$ & $1$ & $1$ & $u_1=(q_1 g_1 \bar{q}_2 q_2 \bar{q}_1)$, $v=(g_2)$ \\
\hline
 $4$ & $\delta_{i_{q_1} j_{\bar{q}_2}} \delta_{i_{q_2} j_{\bar{q}_1}} \delta_{i_{g_1} j_{g_2}} \delta_{i_{g_2} j_{g_1}}$ & $1$ & $2$ & $u_1=(q_1 \bar{q}_2 q_2 \bar{q}_1)$, $v=(g_1 g_2)$ \\
\hline
 $5$ & $N \; \delta_{i_{q_1} j_{g_1}} \delta_{i_{g_1} j_{g_2}} \delta_{i_{g_2} j_{\bar{q}_1}} \delta_{i_{q_2} j_{\bar{q}_2}}$ & $2$ & $0$ & $u_1=(q_1 g_1 g_2 \bar{q}_1)$, $u_2=(q_2 \bar{q}_2)$, $v=()$ \\
\hline
 $6$ & $N \; \delta_{i_{q_1} j_{g_1}} \delta_{i_{g_1} j_{\bar{q}_1}} \delta_{i_{q_2} j_{g_2}} \delta_{i_{g_2} j_{\bar{q}_2}}$ & $2$ & $0$ & $u_1=(q_1 g_1 \bar{q}_1)$, $u_2=(q_2 g_2 \bar{q}_2)$, $v=()$ \\
\hline
 $7$ & $\delta_{i_{q_1} j_{g_1}} \delta_{i_{g_1} j_{\bar{q}_1}} \delta_{i_{q_2} j_{\bar{q}_2}} \delta_{i_{g_2} j_{g_2}}$ & $2$ & $1$ & $u_1=(q_1 g_1 \bar{q}_1)$, $u_2=(q_2 \bar{q}_2)$, $v=(g_2)$ \\
\hline
 $8$ & $\delta_{i_{q_1} j_{\bar{q}_1}} \delta_{i_{q_2} j_{\bar{q}_2}} \delta_{i_{g_1} j_{g_2}} \delta_{i_{g_2} j_{g_1}}$ & $2$ & $2$ & $u_1=(q_1 \bar{q}_1)$, $u_2=(q_2 \bar{q}_2)$, $v=(g_1 g_2)$ \\
\hline
\end{tabular}
\caption{\label{table_example}
The inequivalent colour factors for the amplitude ${\cal A}^{(1)}_6(q_1,\bar{q}_1,q_2,\bar{q}_2,g_1,g_2)$, together with the number of cycles $r$ of the permutation $\pi$,
the number $m$ of gluons attached to the closed string
and the translation to cyclic words.
}
\end{center}
\end{table}
Table~\ref{table_example} shows the inequivalent colour factors for the decomposition of the full amplitude into partial amplitudes.
All other colour factors can be obtained by a re-naming of particles.
There are eight inequivalent colour factors.
Also shown in table~\ref{table_example} is the number of cycles $r$ of the permutation $\pi$.
With two quark-antiquark lines $\pi$ can either be
\bq
 \pi = \left(1, 2\right) 
 & \mbox{or} & 
 \pi = \left(1\right) \left(2\right).
\eq
The first case corresponds to $r=1$, the second case to $r=2$.
In addition we show the number $m$ of gluons attached to the closed string and the translation of the colour factors into cyclic words.

The cases $3$ and $7$ have $m=1$ and are not relevant for NLO calculations: The colour factor will give a vanishing contribution when
contracted into the colour projector of eq.~(\ref{projection_operator}).
In a colour trace basis these cases correspond to
\bq
 \mathrm{Tr}\left(T^{a_{g_2}}\right) & = & 0.
\eq
In the following we will discuss two specific colour factors (case $1$ and case $8$) in detail.

\subsubsection{The first colour factor}

We first discuss the case $1$.
Here we have $r=1$, $m=0$ and the cyclic words
\bq
 u_1 = \left(q_1 g_1 g_2 \bar{q}_2 q_2 \bar{q}_1 \right), & & v=().
\eq
The empty word $v=()$ contributes a factor $N$ to the colour factor.
We start with the non-$N_f$ contribution.
The first line of eq.~(\ref{decomposition}) reduces to a single term and gives the contribution
\bq
 P^{(1) [1]}_{6}\left(q_1^L g_1 g_2 \bar{q}_2^L q_2^L \bar{q}_1^L\right).
\eq
We recall that in $u_1^{[1]}$ all quark lines are assigned a left-moving routing label.
The second line of eq.~(\ref{decomposition}) requires $r>1$ and gives therefore in this case no contribution.
For the third line of eq.~(\ref{decomposition}) we have to compute $CU(u_1)$.
This is done according to algorithm 2 of sect.~\ref{subsect:loop_closing_3}. We first find
\bq
 CU\left(u_1\right)
 & = &
 C_{11}\left(u_1\right)
 +
 C_{22}\left(u_1\right)
 +
 C_{12}\left(u_1\right).
\eq
The loop closing operations on a single quark line yield
\bq
 C_{11}\left(u_1\right) & = & \left(q_1^R g_1 g_2 \bar{q}_2^L q_2^L \bar{q}_1^R \right),
 \nonumber \\
 C_{22}\left(u_1\right) & = & \left(q_1^L g_1 g_2 \bar{q}_2^R q_2^R \bar{q}_1^L \right).
\eq
For the loop closing operation between two different quark lines we have first
\bq
 C_{12}\left(u_1\right)
 & = &
 C_{12}^{LL}\left(u_1\right)
 +
 C_{12}^{LR}\left(u_1\right)
 +
 C_{12}^{RL}\left(u_1\right)
 +
 C_{12}^{RR}\left(u_1\right).
\eq
For the individual terms we find
\bq
 C_{12}^{LL}\left(u_1\right) & = &
 \left( q_1^L g_1 g_2 \bar{q}_2^L q_2^L \bar{q}_1^L \right),
 \nonumber \\
 C_{12}^{LR}\left(u_1\right) & = &
 \left( q_1^L g_1 g_2 q_2^R \bar{q}_2^R \bar{q}_1^L \right)
 +
 \left( q_1^L g_1 q_2^R g_2 \bar{q}_2^R \bar{q}_1^L \right)
 +
 \left( q_1^L q_2^R g_1 g_2 \bar{q}_2^R \bar{q}_1^L \right),
 \nonumber \\
 C_{12}^{RL}\left(u_1\right) & = &
 \left( q_1^R g_1 g_2 \bar{q}_1^R \bar{q}_2^L q_2^L \right)
 +
 \left( q_1^R g_1 \bar{q}_1^R g_2 \bar{q}_2^L q_2^L \right)
 +
 \left( q_1^R \bar{q}_1^R g_1 g_2 \bar{q}_2^L q_2^L \right),
 \nonumber \\
 C_{12}^{RR}\left(u_1\right) & = &
 \left( q_1^R g_1 g_2 \bar{q}_1^R q_2^R \bar{q}_2^R \right)
 +
 \left( q_1^R g_1 \bar{q}_1^R g_2 q_2^R \bar{q}_2^R \right)
 +
 \left( q_1^R g_1 \bar{q}_1^R q_2^R g_2 \bar{q}_2^R \right)
 \nonumber \\
 & &
 +
 \left( q_1^R \bar{q}_1^R g_1 g_2 q_2^R \bar{q}_2^R \right)
 +
 \left( q_1^R \bar{q}_1^R g_1 q_2^R g_2 \bar{q}_2^R \right)
 +
 \left( q_1^R \bar{q}_1^R q_2^R g_1 g_2 \bar{q}_2^R \right).
\eq
Putting everything together and using the reflection identity in eq.~(\ref{reflection_identity})
we obtain for the non-$N_f$ part:
\bq
\lefteqn{
 A^{(1) [1]}_{6,0}
 = } & &
 \nonumber \\
 & &
 \left( 1 - \frac{1}{N^2} \right)
 P^{(1) [1]}_{6}\left(q_1^L g_1 g_2 \bar{q}_2^L q_2^L \bar{q}_1^L\right)
 - \frac{1}{N^2}
 \left[
 P^{(1) [1]}_{6}\left(q_1^L \bar{q}_1^L q_2^R \bar{q}_2^R g_2 g_1 \right)
 +
 P^{(1) [1]}_{6}\left(q_1^L g_1 g_2 \bar{q}_2^R q_2^R \bar{q}_1^L \right)
 \right. 
 \nonumber \\
 & &
 \left.
 +
 P^{(1) [1]}_{6}\left( q_1^L g_1 g_2 q_2^R \bar{q}_2^R \bar{q}_1^L \right)
 +
 P^{(1) [1]}_{6}\left( q_1^L g_1 q_2^R g_2 \bar{q}_2^R \bar{q}_1^L \right)
 +
 P^{(1) [1]}_{6}\left( q_1^L q_2^R g_1 g_2 \bar{q}_2^R \bar{q}_1^L \right)
 \right.
 \nonumber \\
 & &
 \left.
 +
 P^{(1) [1]}_{6}\left( q_1^L q_2^R \bar{q}_2^R \bar{q}_1^L g_2 g_1 \right)
 +
 P^{(1) [1]}_{6}\left( q_1^L q_2^R \bar{q}_2^R g_2 \bar{q}_1^L g_1 \right)
 +
 P^{(1) [1]}_{6}\left( q_1^L q_2^R \bar{q}_2^R g_2 g_1 \bar{q}_1^L \right)
 \right.
 \nonumber \\
 & &
 \left.
 +
 P^{(1) [1]}_{6}\left( q_1^L \bar{q}_2^L q_2^L \bar{q}_1^L g_2 g_1 \right)
 +
 P^{(1) [1]}_{6}\left( q_1^L \bar{q}_2^L q_2^L g_2 \bar{q}_1^L g_1 \right)
 +
 P^{(1) [1]}_{6}\left( q_1^L \bar{q}_2^L g_2 q_2^L \bar{q}_1^L g_1 \right)
 \right.
 \nonumber \\
 & &
 \left.
 +
 P^{(1) [1]}_{6}\left( q_1^L \bar{q}_2^L q_2^L g_2 g_1 \bar{q}_1^L \right)
 +
 P^{(1) [1]}_{6}\left( q_1^L \bar{q}_2^L g_2 q_2^L g_1 \bar{q}_1^L \right)
 +
 P^{(1) [1]}_{6}\left( q_1^L \bar{q}_2^L g_2 g_1 q_2^L \bar{q}_1^L \right)
 \right].
\eq
This agrees with \cite{Ita:2011ar}.

Let us now consider the $N_f$-contribution.
The first line of eq.~(\ref{nf_decomposition}) gives the term
\bq
 \frac{N_f}{N}
 P^{(1) [1/2]}_{6}\left(q_1^L g_1 g_2 \bar{q}_2^L q_2^L \bar{q}_1^L\right).
\eq
For the second line of eq.~(\ref{nf_decomposition}) we would have to compute
\bq
 U\left( u_1, v \right)
 & = &
 \left( \bar{q}_1^L q_2^R \bar{q}_2^R g_2 g_1 q_1^L \right)
 +
 \left( \bar{q}_2^L g_2 g_1 q_1 \bar{q}_1^R q_2^L \right).
\eq
However, these terms correspond to tadpoles and vanish in a massless theory within dimensional regularisation.
Ignoring the tadpoles we 
obtain for the $N_f$-contribution
\bq
 A^{(1) [1/2]}_{6,0}
 & = &
 \frac{N_f}{N}
  P^{(1) [1/2]}_{6}\left(q_1^L g_1 g_2 \bar{q}_2^L q_2^L \bar{q}_1^L\right),
\eq
again in agreement with \cite{Ita:2011ar}.

\subsubsection{The last colour factor}

As a further example we consider the case $8$.
This example will show that we do not necessarily obtain the shortest decomposition of partial amplitudes into primitive amplitudes.
In the case $8$ we have $r=2$, $m=2$ and the cyclic words
\bq
 u_1 = \left( q_1 \bar{q}_1 \right), 
 \;\;\;
 u_2 = \left( q_2 \bar{q}_2 \right), 
 \;\;\;
 v = \left( g_1 g_2 \right).
\eq
In the non-$N_f$-part only the first line of eq.~(\ref{decomposition}) contributes due to $m=2$.
We have
\bq
\label{eq_example_cyclic_shuffle}
 U\left( u_1^{[1]}, u_2 \right) \circledcirc v^T
 & = &
 \left( q_1^L \bar{q}_1^L q_2^R \bar{q}_2^R \right) \circledcirc \left( g_2 g_1 \right)
 \nonumber \\
 U\left( u_1, u_2^{[1]} \right) \circledcirc v^T
 & = &
 \left( q_1^R \bar{q}_1^R q_2^L \bar{q}_2^L \right) \circledcirc \left( g_2 g_1 \right)
\eq
Working out the cyclic shuffle product is not too complicated and will produce in each case an expression with twenty terms.
We obtain
\bq
 A^{(1) [1]}_{6,2}
 & = &
 - \frac{1}{N} \left[
 P^{(1) [1]}_{6}\left( q_1^L \bar{q}_1^L q_2^R \bar{q}_2^R g_2 g_1 \right)
 +
 P^{(1) [1]}_{6}\left( q_1^L \bar{q}_1^L q_2^R g_2 \bar{q}_2^R g_1 \right)
 +
 P^{(1) [1]}_{6}\left( q_1^L \bar{q}_1^L g_2 q_2^R \bar{q}_2^R g_1 \right)
 \right.
 \nonumber \\
 & &
 \left.
 +
 P^{(1) [1]}_{6}\left( q_1^L g_2 \bar{q}_1^L q_2^R \bar{q}_2^R g_1 \right)
 +
 P^{(1) [1]}_{6}\left( q_1^L \bar{q}_1^L q_2^R g_2 g_1 \bar{q}_2^R \right)
 +
 P^{(1) [1]}_{6}\left( q_1^L \bar{q}_1^L g_2 q_2^R g_1 \bar{q}_2^R \right)
 \right.
 \nonumber \\
 & &
 \left.
 +
 P^{(1) [1]}_{6}\left( q_1^L g_2 \bar{q}_1^L q_2^R g_1 \bar{q}_2^R \right)
 +
 P^{(1) [1]}_{6}\left( q_1^L \bar{q}_1^L g_2 g_1 q_2^R \bar{q}_2^R \right)
 +
 P^{(1) [1]}_{6}\left( q_1^L g_2 \bar{q}_1^L g_1 q_2^R \bar{q}_2^R \right)
 \right.
 \nonumber \\
 & &
 \left.
 +
 P^{(1) [1]}_{6}\left( q_1^L g_2 g_1 \bar{q}_1^L q_2^R \bar{q}_2^R \right)
 \right]
 +
 \left[ g_1 \leftrightarrow g_2 \right]
 +
 \left[ L \leftrightarrow R \right].
\eq
The operation $[ g_1 \leftrightarrow g_2 ]$ instructs us to add terms with $g_1$ and $g_2$ exchanged,
the operation $[ L \leftrightarrow R ]$ instructs us to add on top of that terms with all routing labels exchanged.

For the $N_f$-part we have to consider
\bq
\label{eq_U_operation}
 U\left( u_1, u_2, v \right)
  = 
 \left( \bar{q}_1^L q_1^L \right) \circledcirc \left( \bar{q}_2^L q_2^L \right) \circledcirc \left( g_1 g_2 \right)
 +
 \left( q_1^L \bar{q}_1^L q_2^R \bar{q}_2^R \right) \circledcirc \left( g_1 g_2 \right)
 +
 \left( q_1^R \bar{q}_1^R q_2^L \bar{q}_2^L \right) \circledcirc \left( g_1 g_2 \right).
 \nonumber \\
\eq
The last two terms generate expressions identical to the one of eq.~(\ref{eq_example_cyclic_shuffle}).
For the first term we start with
\bq
\label{eq_example_cyclic_shuffle_2}
 \left( \bar{q}_1^L q_1^L \right) \circledcirc \left( \bar{q}_2^L q_2^L \right) 
 & = &
 \left( \bar{q}_1^L q_1^L \bar{q}_2^L q_2^L \right)
 +
 \left( \bar{q}_1^L \bar{q}_2^L q_2^L q_1^L \right)
 +
 \left( \bar{q}_1^L q_1^L q_2^L \bar{q}_2^L \right).
\eq
The cyclic order $(\bar{q}_1^L q_2^L \bar{q}_2^L q_1^L)$ cannot be drawn without crossed fermion lines and is therefore not included.
The cyclic shuffle product of each of the three terms in eq.~(\ref{eq_example_cyclic_shuffle_2}) with $(g_1 g_2)$ will generate twenty terms.
However, the terms coming from the last two terms in eq.~(\ref{eq_example_cyclic_shuffle_2}) will cancel with the terms coming from the last two terms
of eq.~(\ref{eq_U_operation}).
Thus only the terms coming from the first term of eq.~(\ref{eq_example_cyclic_shuffle_2}) survive.
Putting everything together we obtain for the $N_f$-part
\bq
\lefteqn{
 A^{(1) [1/2]}_{6,2}
 = 
  \frac{N_f}{N^2}
 } & & 
 \nonumber \\
 & &
 \left[
 P^{(1) [1/2]}_{6}\left( q_1^L \bar{q}_2^L q_2^L \bar{q}_1^L g_1 g_2 \right)
 +
 P^{(1) [1/2]}_{6}\left( q_1^L \bar{q}_2^L q_2^L g_1 \bar{q}_1^L g_2 \right)
 +
 P^{(1) [1/2]}_{6}\left( q_1^L \bar{q}_2^L g_1 q_2^L \bar{q}_1^L g_2 \right)
 \right. \nonumber \\
 & & \left.
 +
 P^{(1) [1/2]}_{6}\left( q_1^L g_1 \bar{q}_2^L q_2^L \bar{q}_1^L g_2 \right)
 +
 P^{(1) [1/2]}_{6}\left( q_1^L \bar{q}_2^L q_2^L g_1 g_2 \bar{q}_1^L \right)
 +
 P^{(1) [1/2]}_{6}\left( q_1^L \bar{q}_2^L g_1 q_2^L g_2 \bar{q}_1^L \right)
 \right. \nonumber \\
 & & \left.
 +
 P^{(1) [1/2]}_{6}\left( q_1^L g_1 \bar{q}_2^L q_2^L g_2 \bar{q}_1^L \right)
 +
 P^{(1) [1/2]}_{6}\left( q_1^L \bar{q}_2^L g_1 g_2 q_2^L \bar{q}_1^L \right)
 +
 P^{(1) [1/2]}_{6}\left( q_1^L g_1 \bar{q}_2^L g_2 q_2^L \bar{q}_1^L \right)
 \right. \nonumber \\
 & & \left.
 +
 P^{(1) [1/2]}_{6}\left( q_1^L g_1 g_2 \bar{q}_2^L q_2^L \bar{q}_1^L \right)
 \right]
 +
 \left[ g_1 \leftrightarrow g_2 \right].
\eq
We remark that shuffle operations will give a decomposition of partial amplitudes into primitive amplitudes with a high degree of symmetry,
but not necessarily with a minimum number of terms.
Shorter expressions are possible, as can be seen by comparing our result with the result of \cite{Ita:2011ar}.

The result of the decomposition based on shuffle relations can always be simplified with respect to known side relations.
Known side relations are for example trivial fermionic flip identities like in eq.~(\ref{side_relation}) and the reflection identity in eq.~(\ref{reflection_identity}),
relations following from Furry's theorem as discussed in \cite{Ellis:2011cr,Badger:2012pg},
or relations obtained by exploiting the freedom of choice in the shuffle operations as discussed in section~\ref{sect:operations}.
Obtaining a decomposition of the partial amplitudes into primitive amplitudes with a minimal number of terms is of interest if the physical
observable is computed as the sum over all contributing primitive amplitudes.
However, also the minimal set of independent primitive amplitudes grows quite fast with the number of external particles and it can be advantageous
to replace the sum over all contributing primitive amplitudes by a Monte Carlo sampling over the primitive amplitudes.
In this case the absolute number of contributing primitive amplitudes is not so important, what matters more is the mapping from a set of random numbers
to the primitive amplitudes.

\section{Conclusions}
\label{sect:conclusions}

In this paper we presented a method for the decomposition of QCD partial amplitudes into primitive amplitudes at one-loop level and tree level
for arbitrary numbers of quarks and gluons.
Our method is entirely combinatorial and based on shuffle relations.
The method avoids Feynman diagrams and does not require to solve a system of linear equations.
In many cases of interest our method allows the possibility that the decomposition into primitive amplitudes can be worked out
at the back of an envelope, whereas with other methods one might already have to resort to computers.
Of course, our method is also suited for an automated computer implementation.
The techniques and methods presented in this paper will be useful for the computation of one-loop amplitudes with many external legs.
Shuffle products are easily computed and for all practical applications the decomposition of the partial amplitudes into primitive amplitudes
will not be the bottle-neck.

\subsection*{Acknowledgements}

The work of C.R. is supported in part by the BMBF.
S.W. would like to thank the Kavli Institute for Theoretical Physics in Santa Barbara 
and the Simons Center for Geometry and Physics in Stony Brook for hospitality, where part of this project was carried out.
This research was supported in part by the National Science Foundation under Grant No. NSF PHY11-25915.


\begin{appendix}


\section{Feynman rules}
\label{appendix:colour_ordered_rules}

In this appendix we give a list of the colour ordered Feynman rules. 
They are obtained from the standard Feynman rules by extracting from each formula the coupling
constant and the colour part.
The propagators for quark, gluon and ghost particles are given by
\bq
\begin{picture}(85,20)(0,5)
 \ArrowLine(70,10)(20,10)
\end{picture} 
 & = &
 i\frac{k\!\!\!/+m}{k^2-m^2},
 \nonumber \\
\begin{picture}(85,20)(0,5)
 \Gluon(20,10)(70,10){-5}{5}
\end{picture} 
& = &
 \frac{-ig^{\mu\nu}}{k^2},
 \nonumber \\
\begin{picture}(85,20)(0,5)
 \DashArrowLine(70,10)(20,10){3}
\end{picture} 
 & = &
 \frac{i}{k^2}.
\eq
The colour ordered Feynman rules for the three-gluon and the four-gluon vertices are
\bq
\begin{picture}(100,35)(0,50)
\Vertex(50,50){2}
\Gluon(50,50)(50,80){3}{4}
\Gluon(50,50)(76,35){3}{4}
\Gluon(50,50)(24,35){3}{4}
\LongArrow(56,70)(56,80)
\LongArrow(67,47)(76,42)
\LongArrow(33,47)(24,42)
\Text(60,80)[lt]{$k_1^{\mu_1}$}
\Text(78,35)[lc]{$k_2^{\mu_2}$}
\Text(22,35)[rc]{$k_3^{\mu_3}$}
\end{picture}
 & = &
 i \left[ g^{\mu_1\mu_2} \left( k_2^{\mu_3} - k_1^{\mu_3} \right)
         +g^{\mu_2\mu_3} \left( k_3^{\mu_1} - k_2^{\mu_1} \right)
         +g^{\mu_3\mu_1} \left( k_1^{\mu_2} - k_3^{\mu_2} \right)
   \right],
 \nonumber \\
 \nonumber \\
 \nonumber \\
\begin{picture}(100,35)(0,50)
\Vertex(50,50){2}
\Gluon(50,50)(71,71){3}{4}
\Gluon(50,50)(71,29){3}{4}
\Gluon(50,50)(29,29){3}{4}
\Gluon(50,50)(29,71){3}{4}
\Text(72,72)[lb]{\small $\mu_1$}
\Text(72,28)[lt]{\small $\mu_2$}
\Text(28,28)[rt]{\small $\mu_3$}
\Text(28,72)[rb]{\small $\mu_4$}
\end{picture}
 & = &
  i \left[
        2 g^{\mu_1\mu_3} g^{\mu_2\mu_4} - g^{\mu_1\mu_2} g^{\mu_3\mu_4} 
                                        - g^{\mu_1\mu_4} g^{\mu_2\mu_3}
 \right].
 \nonumber \\
 \nonumber \\
\eq
The colour-ordered three-gluon vertex is anti-symmetric under the exchange of two external legs.
The associated colour factor is also anti-symmetric under the exchange of two external legs.
The Feynman rule for the full three-gluon vertex (including colour) is therefore symmetric 
under the exchange of two external legs, as it should be.

We can eliminate the four-gluon vertex \cite{Draggiotis:1998gr,Duhr:2006iq} by introducing a tensor particle with propagator
\bq
\begin{picture}(85,20)(0,5)
 \Line(70,12)(20,12)
 \Line(70,8)(20,8)
\Text(18,12)[rb]{\small $\alpha_1$}
\Text(72,12)[lb]{\small $\alpha_2$}
\Text(18,8)[rt]{\small $\beta_1$}
\Text(72,8)[lt]{\small $\beta_2$}
\end{picture} 
 & = &
 -\frac{i}{2} g^{\alpha_1\alpha_2} g^{\beta_1\beta_2}
 \nonumber \\
\eq
and a gluon-gluon-tensor vertex
\bq
\begin{picture}(100,35)(0,50)
\Vertex(50,50){2}
\Gluon(50,50)(29,29){3}{4}
\Gluon(50,50)(29,71){3}{4}
\Line(50,52)(81,52)
\Line(50,48)(81,48)
\Text(28,28)[rt]{\small $\mu_1$}
\Text(28,72)[rb]{\small $\mu_2$}
\Text(83,52)[lb]{\small $\alpha$}
\Text(83,48)[lt]{\small $\beta$}
\end{picture}
 & = &
 i \left( g^{\mu_1\alpha} g^{\mu_2\beta} - g^{\mu_1\beta} g^{\mu_2\alpha} \right).
 \nonumber \\
 \nonumber \\
\eq
The gluon-gluon-tensor vertex is anti-symmetric under the exchange of the two gluons.

We turn to the ghost-antighost-gluon vertex.
We have now a three-valent vertex with three distinguishable particles.
The colour factor ($i f^{abc}$) is again anti-symmetric under the exchange of two external legs.
Since the particles are distinguishable, we now have to distinguish between the two inequivalent cyclic orderings
ghost-gluon-antighost and gluon-ghost-antighost.
Hence, the colour-ordered Feynman rules are given by
\bq
\begin{picture}(100,35)(0,50)
\Vertex(50,50){2}
\Gluon(50,50)(80,50){3}{4}
\DashArrowLine(50,50)(29,71){3}
\DashArrowLine(29,29)(50,50){3}
\LongArrow(36,59)(29,66)
\Text(82,50)[lc]{$\mu$}
\Text(28,71)[rb]{$k$}
\end{picture}
 \;\; = \;\;
 -i k^{\mu},
 & \;\;\;\;\;\;\;\;\; &
\begin{picture}(100,35)(0,50)
\Vertex(50,50){2}
\Gluon(50,50)(20,50){3}{4}
\DashArrowLine(50,50)(71,71){3}
\DashArrowLine(71,29)(50,50){3}
\LongArrow(64,59)(71,66)
\Text(18,50)[rc]{$\mu$}
\Text(72,71)[lb]{$k$}
\end{picture}
 \;\; = \;\;
 i k^{\mu}.
 \nonumber \\
 \nonumber \\
\eq
We finally discuss the quark-antiquark-gluon vertex.
Suppose the quarks are in the adjoint representation.
Then the full Feynman rules for the quark-antiquark-gluon vertex
is $g f^{a_qb_gc_{\bar{q}}} \gamma^\mu$
and we obtain the colour-ordered Feynman rules
\bq
\label{colour_ordered_quark_gluon_antiquark_vertex}
\begin{picture}(100,35)(0,50)
\Vertex(50,50){2}
\Gluon(50,50)(80,50){3}{4}
\ArrowLine(50,50)(29,71)
\ArrowLine(29,29)(50,50)
\Text(82,50)[lc]{$\mu$}
\end{picture}
 \;\; = \;\;
 -i \gamma^{\mu},
 & \;\;\;\;\;\;\;\;\; &
\begin{picture}(100,35)(0,50)
\Vertex(50,50){2}
\Gluon(50,50)(20,50){3}{4}
\ArrowLine(50,50)(71,71)
\ArrowLine(71,29)(50,50)
\Text(18,50)[rc]{$\mu$}
\end{picture}
 \;\; = \;\;
 i \gamma^{\mu}.
 \nonumber \\
 \nonumber \\
\eq
These Feynman rules are again anti-symmetric under the exchange of two external legs.
It is advantageous to have all colour-ordered Feynman rules for three-valent vertices to be anti-symmetric
under the exchange of two particles.
In the case where the quarks are in the fundamental representation we define therefore the colour-ordered
quark-gluon-antiquark vertex to be given by eq.~(\ref{colour_ordered_quark_gluon_antiquark_vertex}).
The associated colour factors are then $T^a_{ij}$ for the cyclic ordering quark-gluon-antiquark
and $(-T^a_{ij})$ for the cyclic ordering gluon-quark-antiquark.
The full Feynman rules (including  colour) is in all cases $-i g T^a_{ij} \gamma^\mu$.

\section{The shuffle product for $U(1)$-gluons}
\label{appendix:U1_shuffle}

In this appendix we show that the operation $U_{ij}$
corresponds to the case where two quark lines $i$ and $j$ are connected through a $U(1)$-gluon.
We let
\bq
 u = \left( l_1 ... l_k \right) = \left( \bar{q}_i u_{i,L} q_i u_{i,R} \right),
 & &
 v = \left( l_{k+1} ... l_r \right) = \left( \bar{q}_j v_{j,L} q_j v_{j,R} \right),
\eq
where $u_{i,L}$, $u_{i,R}$, $v_{j,L}$ and $v_{j,R}$ are words from the alphabet
$A=\{\bar{q}_1,\bar{q}_2,...,q_1,q_2,...,g_1,g_2,...\}$ not containing the letters $\bar{q}_i$, $q_i$, $\bar{q}_j$ and $q_j$.
The operation $U_{ij}$ is defined by
\bq
 U_{ij}\left( u, v \right)
 & = &
 \sum\limits_{\stackrel{(\mbox{\tiny cyclic shuffles} \; \sigma) / {\mathbb Z}_r}{(q_i ... \bar{q}_i ... q_j ... \bar{q}_j ...)}} \left( l_{\sigma(1)} l_{\sigma(2)} ... l_{\sigma(r)} \right),
\eq
where the sum is over all cyclic shuffles with the cyclic order of the quarks/antiquarks given by $(q_i ... \bar{q}_i ... q_j ... \bar{q}_j ... )$.
We consider
\bq
 \sum\limits_{w \in U_{ij}\left(u,v\right)} P_r^{(0)}(w)
\eq
and show that in the sum only the diagrams survive, which correspond to the diagrams of the primitive amplitudes 
$P_k^{(0)}(u)$ and $P_{r-k}^{(0)}(v)$ connected by a $U(1)$-gluon between the two quark lines $i$ and $j$.

We show first, that all other diagrams drop out. There is a straightforward proof, which is however tedious and notational cumbersome.
The following version is shorter and more elegant: The operation $U_{ij}$ is commutative and we have a problem symmetric
in $i$ and $j$.
In order to show that all other diagrams drop out, it is therefore sufficient to show that
\begin{description}
\item{(a)} The quark lines $i$ and $j$ are connected by a single gluon with no side branchings.
\item{(b)} Apart from that gluon only currents from the word $v$ are connected to the quark line $j$.
\end{description}
By the symmetry of the problem it follows then that the quark line $i$ is only connected to currents from the word $u$ and the single
gluon mentioned in $(a)$.
Note that in $(b)$ we don't have to show that all particles from $v$ are connected to the quark line $j$ through currents,
it is sufficient to show that no other particles are directly connected to the quark line $j$.

Let
\bq
 v_{j,L} = v_1 v_2 v_3
\eq
be a decomposition of the word $v_{j,L}$, such that in the shuffle product $v_1$ appears between $\bar{q}_j$ and $q_i$,
$v_2$ appears between $q_i$ and $\bar{q}_i$ and $v_3$ appears between $\bar{q}_i$ and $q_j$.
For this decomposition we have on the right of the quark line $i$ all shuffles $v_2 \Sha u_{i,R}$, while on the left of the quark
line $i$ we have all shuffles $u_{i,L} \Sha v_3 q_j v_{j,R} \bar{q}_j v_1$.
We can now use the same argument as in the proof of the Kleiss-Kuijf relation and deduce that the currents connected to the 
left of the quark line $i$ are either made out of particles from $u_{i,L}$ or from particles from $v_3 q_j v_{j,R} \bar{q}_j v_1$, but not mixed.
This shows that the quark lines $i$ and $j$ are connected by a gluon with possible side-branchings with particles from $v$, and no other
currents from $u$ are connected to the quark line $j$. 
The gluon connecting the quark lines $i$ and $j$ has no side-branchings with particles from $u$, exchanging the roles of $i$ and $j$
shows that it cannot have side-branchings with particles from $v$. Therefore this gluon has no side-branchings at all
and is therefore a $U(1)$-gluon. This proves $(a)$ and $(b)$.

To complete the proof one verifies that all diagrams corresponding to $U(1)$-gluons are generated by the shuffle product $U_{ij}$
and not cancelled by other contributions.
This part is trivial and left to the reader.

\section{The loop closing operation}
\label{appendix:loop_closing}

In this appendix we outline a proof, that the operations $C_{ii}(u)$ -- defined in eq.~(\ref{U1_closing_ii}) --
and $C_{ij}(u)$ -- defined through eqs.~(\ref{def_C_ij})-(\ref{def_C_ij_LL}) -- correspond to the loop closing operation with a $U(1)$-gluon.
We start with the operation $C_{ii}(u)$, where the cyclic word is given by
\bq
 u & = & 
 \left( q_i u_1 \bar{q}_i u_2 \right).
\eq
Before the loop closing operation we can think of the word $u$ as defining a primitive tree amplitude.
Closing the loop with a $U(1)$-gluon on the quark line $i$ means to consider all diagrams which are obtained from the diagrams
of the primitive tree amplitude by emitting and re-absorbing a $U(1)$-gauge boson from the quark line $i$ in all possible ways.
Our task is to re-express this set of diagrams as a linear combination of primitive one-loop amplitudes.
Without loss of generality we can draw all diagrams such that the $U(1)$-gluon is emitted to the left of the quark line $i$.
With this convention the problem is trivial if the word $u_2$ is empty:
In this case all other particles are attached to the right side of the quark line $i$ and only the $U(1)$-gluon couples from the left side of the 
quark line $i$.
In this case the linear combination of primitive one-loop amplitudes consists of a single term, given by the primitive one-loop amplitude
with cyclic order $(q_i u_1 \bar{q}_i)$.

In the case, where $u_2$ is not empty we reduce the problem to the trivial case discussed above.
This is done as follows: The $U(1)$-gluon on the left side of the quark line $i$ does not interact with the particles from the cyclic word $u_2$
(also on the left side of the quark line $i$).
We can flip all particles from $u_2$ to the other side of the quark line $i$, this gives the sign
\bq
 \left(-1\right)^{n_{u_2}}.
\eq
The particles of $u_2$ are then on the right side of the quark line $i$ in the reversed order. 
There they do not interact with the particles from $u_1$.
The combination 
\bq
 u_1 \Sha u_2^T 
\eq
gives the correct linear combination of cyclic orders, which describe this situation.
This is again the Kleiss-Kuijf relation.
Therefore we have reduced the problem to the case, where only the $U(1)$-gluon couples from the left side to the quark line $i$.

We now turn to the operation $C_{ij}(u)$, where a $U(1)$-gluon closes the loop between quark lines $i$ and $j$.
\begin{figure}
\begin{center}
\begin{picture}(110,100)(0,0)
\Vertex(20,50){2}
\Vertex(80,50){2}
\Gluon(20,50)(80,50){6}{5}
\ArrowLine(10,90)(20,50)
\ArrowLine(20,50)(10,10)
\ArrowLine(90,10)(80,50)
\ArrowLine(80,50)(90,90)
\Vertex(12.5,20){2}
\Vertex(87.5,20){2}
\Photon(12.5,20)(87.5,20){3}{8}
\Text(50,10)[t]{$RR$}
\Text(5,10)[r]{\small $q_i$}
\Text(5,90)[r]{\small $\bar{q}_i$}
\Text(95,90)[l]{\small $q_j$}
\Text(95,10)[l]{\small $\bar{q}_j$}
\end{picture} 
\begin{picture}(110,100)(0,0)
\Vertex(20,50){2}
\Vertex(80,50){2}
\Gluon(20,50)(80,50){6}{5}
\ArrowLine(10,90)(20,50)
\ArrowLine(20,50)(10,10)
\ArrowLine(90,10)(80,50)
\ArrowLine(80,50)(90,90)
\Vertex(12.5,20){2}
\Vertex(87.5,80){2}
\PhotonArc(12.5,50)(30,270,340){3}{5}
\PhotonArc(87.5,50)(30,90,160){3}{5}
\Text(50,10)[t]{$RL$}
\end{picture} 
\begin{picture}(110,100)(0,0)
\Vertex(20,50){2}
\Vertex(80,50){2}
\Gluon(20,50)(80,50){6}{5}
\ArrowLine(10,90)(20,50)
\ArrowLine(20,50)(10,10)
\ArrowLine(90,10)(80,50)
\ArrowLine(80,50)(90,90)
\Vertex(12.5,80){2}
\Vertex(87.5,20){2}
\PhotonArc(12.5,50)(30,20,90){3}{5}
\PhotonArc(87.5,50)(30,200,270){3}{5}
\Text(50,10)[t]{$LR$}
\end{picture} 
\begin{picture}(110,100)(0,0)
\Vertex(20,50){2}
\Vertex(80,50){2}
\Gluon(20,50)(80,50){6}{5}
\ArrowLine(10,90)(20,50)
\ArrowLine(20,50)(10,10)
\ArrowLine(90,10)(80,50)
\ArrowLine(80,50)(90,90)
\Vertex(12.5,80){2}
\Vertex(87.5,80){2}
\Photon(12.5,80)(87.5,80){3}{8}
\Text(50,10)[t]{$LL$}
\end{picture} 
\caption{\label{figure_C_ij}
Attaching a $U(1)$-gluon in all possible between the quark lines $i$ and $j$ of the primitive $(q_i\bar{q}_iq_j\bar{q}_j)$ amplitude.
}
\end{center}
\end{figure}
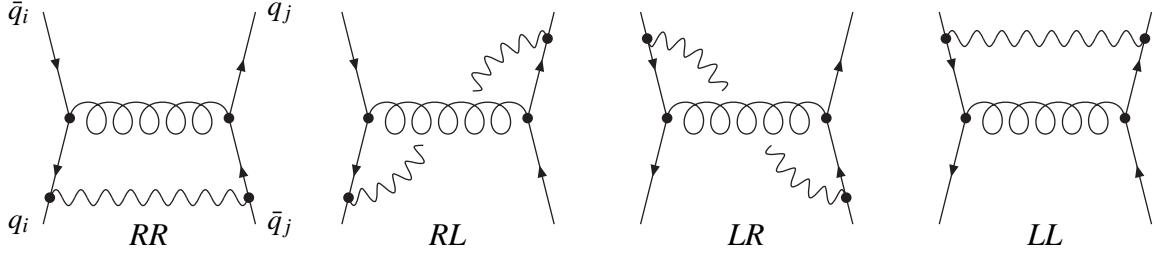
We start from the cyclic word
\bq
 u & = & 
 \left( q_i u_1 \bar{q}_i u_2 q_j u_3 \bar{q}_j u_4 \right).
\eq
The relevant diagrams can be grouped into four categories.
We illustrate the four categories for a simple example in fig.~(\ref{figure_C_ij}).
In the example of fig.~(\ref{figure_C_ij}) the sub-words $u_1$, $u_2$, $u_3$ and $u_4$ are empty.
The four categories correspond to the sum
\bq
 C_{ij}\left(u\right)
 & = &
 C_{ij}^{RR}\left(u\right)
 +
 C_{ij}^{RL}\left(u\right)
 +
 C_{ij}^{LR}\left(u\right)
 +
 C_{ij}^{LL}\left(u\right).
\eq
We split the proof into two parts: We first show that all required diagrams are contained (exactly once) in this sum.
In the second part we show that all undesired diagrams cancel in the sum.
Let us first consider the $RR$-contribution. 
Similar to the discussion for the $C_{ii}$-operation, this case is trivial, if the word $u_4$ is empty.
If $u_4$ is non-empty we can again use the Kleiss-Kuijf relation and flip all particles of $u_4$ above the line $q_i$-$\bar{q}_j$.
We are thus lead to the formula
\bq
 C_{ij}^{RR}\left(u\right)
 & = &
 \left(-1\right)^{n_{u_4}}
 \sum\limits_{w \in \left( q_i^R \left(\left( u_1 \bar{q}_i^R u_2 q_j^R u_3 \right) \Sha u_4^T \right) \bar{q}_j^R \right)}
 w.
\eq
If this formula is promoted to primitive one-loop amplitudes it will
contain all desired diagrams for this case.
However we note that for $C_{ij}$ -- in contrast to the $C_{ii}$-case -- the line connecting $q_i$ with $\bar{q}_j$ is not a continuous quark line.
This implies that the primitive one-loop amplitudes will generate diagrams, where the $U(1)$-gluon couples to these intermediate particles.
In the second step we will show that all these undesired diagrams cancel in the sum.

We continue with the $RL$-contribution.
\begin{figure}
\begin{center}
\bq
\begin{picture}(110,40)(0,40)
\Vertex(20,50){2}
\Vertex(80,50){2}
\Gluon(20,50)(80,50){6}{5}
\ArrowLine(10,90)(20,50)
\ArrowLine(20,50)(10,10)
\ArrowLine(90,10)(80,50)
\ArrowLine(80,50)(90,90)
\Vertex(12.5,20){2}
\Vertex(87.5,80){2}
\PhotonArc(12.5,50)(30,270,340){3}{5}
\PhotonArc(87.5,50)(30,90,160){3}{5}
\Text(5,10)[r]{\small $q_i$}
\Text(5,90)[r]{\small $\bar{q}_i$}
\Text(95,90)[l]{\small $q_j$}
\Text(95,10)[l]{\small $\bar{q}_j$}
\end{picture} 
 & = & 
- 
\begin{picture}(110,40)(0,40)
\Vertex(20,50){2}
\Vertex(80,50){2}
\Gluon(20,50)(80,50){6}{5}
\ArrowLine(10,90)(20,50)
\ArrowLine(20,50)(10,10)
\ArrowLine(80,50)(90,10)
\ArrowLine(70,30)(80,50)
\Vertex(12.5,20){2}
\Vertex(87.5,20){2}
\Photon(12.5,20)(87.5,20){3}{8}
\Text(5,10)[r]{\small $q_i$}
\Text(5,90)[r]{\small $\bar{q}_i$}
\Text(95,10)[l]{\small $q_j$}
\Text(65,30)[r]{\small $\bar{q}_j$}
\end{picture} 
 \nonumber \\
\eq
\caption{\label{figure_C_ij_2}
The $RL$-contribution can be re-drawn in the way shown on the right-hand side. The minus sign is due to the fact, that the $U(1)$-gluon is emitted
now to the right side of the quark line $j$.
}
\end{center}
\end{figure}
We note that diagrams corresponding to this contribution can also be drawn in the way as shown in fig.~(\ref{figure_C_ij_2}).
The minus sign is due to the fact that the $U(1)$-gluon is emitted now to the right side of the quark line $j$.
Again we can use the Kleiss-Kuijf relation to flip all particles below the $q_i$-$q_j$ line to the other side.
The particles below this line are given by the sequence $u_3 \bar{q}_j^L u_4 $ and we recognise in
\bq
 \left(-1\right)^{n_{u_3}+1+n_{u_4}}
 \left( u_1 \bar{q}_i u_2 \right) \Sha \left( u_3 \bar{q}_j u_4 \right)^T 
\eq
the Kleiss-Kuijf relation. 
Flipping $\bar{q}_j$ above will turn the fermion line $\bar{q}_j$-$q_j$ into a left-moving fermion line, which explains the label $RL$.
Including the minus sign from fig.~(\ref{figure_C_ij_2}) we obtain the overall sign as
\bq
 \left(-1\right)^{n_{u_3}+n_{u_4}}.
\eq
This explains eq.~(\ref{def_C_ij_RL}).
The argumentation for the $LR$- and $LL$-contributions is similar and leads to the eqs.~(\ref{def_C_ij_LR}) and~(\ref{def_C_ij_LL}).

We now come to the second part and discuss the cancellation of the undesired diagrams.
\begin{figure}
\begin{center}
\begin{picture}(150,100)(0,0)
\Vertex(20,50){2}
\Vertex(80,50){2}
\Gluon(20,50)(42.5,50){6}{1}
\Vertex(42.5,50){2}
\Gluon(42.5,50)(80,50){6}{3}
\ArrowLine(10,90)(20,50)
\ArrowLine(20,50)(10,10)
\ArrowLine(90,10)(80,50)
\ArrowLine(80,50)(90,90)
\Vertex(12.5,20){2}
\Vertex(87.5,20){2}
\PhotonArc(12.5,50)(30,270,360){3}{5}
\Text(50,10)[t]{$RR$}
\Text(5,10)[r]{\small $q_i$}
\Text(5,90)[r]{\small $\bar{q}_i$}
\Text(95,90)[l]{\small $q_j$}
\Text(95,10)[l]{\small $\bar{q}_j$}
\end{picture} 
\begin{picture}(130,100)(0,0)
\Vertex(20,50){2}
\Vertex(80,50){2}
\Gluon(20,50)(80,50){6}{5}
\Vertex(80,50){2}
\Gluon(80,50)(100,50){6}{1}
\ArrowLine(10,90)(20,50)
\ArrowLine(20,50)(10,10)
\ArrowLine(110,10)(100,50)
\ArrowLine(100,50)(110,90)
\Vertex(12.5,20){2}
\PhotonArc(12.5,50)(30,270,340){3}{5}
\PhotonArc(61,50)(19,0,145){3}{7}
\Text(50,10)[t]{$RL$}
\Text(5,10)[r]{\small $q_i$}
\Text(5,90)[r]{\small $\bar{q}_i$}
\Text(115,90)[l]{\small $q_j$}
\Text(115,10)[l]{\small $\bar{q}_j$}
\end{picture} 
\caption{\label{figure_C_ij_3}
Cancellation of undesired diagrams between the $RR$- and the $RL$-contribution due to the antisymmetry of the vertices.
}
\end{center}
\end{figure}
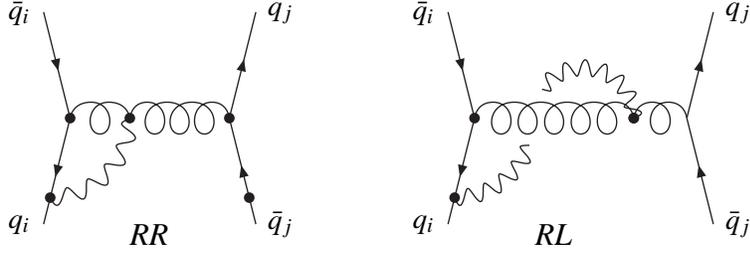
In a primitive amplitude the $U(1)$-gluon looses its abelian character and acts as any other gluon.
As a consequence it couples to other gluons.
Therefore we obtain diagrams as shown in fig.~(\ref{figure_C_ij_3}).
If we sum over all contributions these diagrams cancel due to the antisymmetry of the vertices.
This cancellation is shown for a simple example in fig.~(\ref{figure_C_ij_3}).

\end{appendix}

\bibliography{/home/stefanw/notes/biblio}
\bibliographystyle{/home/stefanw/latex-style/h-physrev5}

\end{document}